\documentclass[twocolumn]{aastex63}

\usepackage{graphicx}
\usepackage{verbatim}
\usepackage{comment}

\usepackage{hyperref}

\usepackage{rotating}
\usepackage{tabularx}

\usepackage{array}

\DeclareUnicodeCharacter{0015}{}

\newcommand{\lgd}{$\Sigma$ }
\newcommand{\rad}{$R$ }

\begin{document}

\title{Pre-processing of Galaxies in the Early Stages of Cluster Formation in Abell 1882 at $z$=0.139}

\author{Aparajita Sengupta}
\affil{Department of Physics, Indiana University-Purdue University Indianapolis, IN 46202, USA}
\author{William C. Keel}
\affiliation{Department of Physics and Astronomy, University of Alabama, Tuscaloosa, AL 35487, USA}
\author{Glenn Morrison}
\affiliation{Institute for Astronomy, University of Hawaii, Honolulu, HI, 96822, USA}
\affiliation{Canada-France-Hawaii Telescope, Kamuela, HI, 96743, USA}
\author{Rogier A. Windhorst}
\affiliation{School of Earth $\&$ Space Exploration, Arizona State University, Tempe, AZ 85287-1404, USA}
\author{Neal Miller}
\affiliation{Department of Mathematics and Physics, Stevenson University, Owings Mills, MD 21117}
\author{Brent Smith}
\affiliation{School of Earth $\&$ Space Exploration, Arizona State University, Tempe, AZ 85287-1404, USA}

\begin{abstract}

A rare opportunity to distinguish internal and environmental effects on galaxy evolution is afforded by ``Supergroups'', systems which are rich and massive but include several comparably rich substructures, surrounded by filaments. We present here a multiwavelength photometric and spectroscopic study of the galaxy population in the Supergroup Abell 1882 at $z$=0.139, combining new data from the MMT and Hectospec with archival results from the Galaxy And Mass Assembly (GAMA) survey, the Sloan Digital Sky Survey (SDSS), NED, the Gemini Multi-Object Spectrograph (GMOS) and Galaxy Evolution Explorer (GALEX). These provide spectroscopic classifications for 526 member galaxies, across wide ranges of local density and velocity dispersion. We identify three prominent filaments along which galaxies seem to be entering the Supergroup (mostly in E-W directions). Abell 1882 has a well-populated red sequence, containing most galaxies with stellar mass $>$ $10^{10.5}$ $M_{sun}$, and a pronounced color-density relation even within its substructures. Thus, galaxy evolution responds to the external environment as strongly in these unrelaxed systems as we find in rich and relaxed clusters. From these data, local density remains the primary factor with a secondary role for distance from the inferred center of the entire structure's potential well. Effects on star formation, as traced by optical and near-UV colors, depend on galaxy mass. We see changes in lower-mass galaxies (M $<$ $10^{10.5}$ $M_{sun}$) at four times the virial radius of major substructures, while the more massive $NUV$ Green Valley galaxies show low levels of star formation within two virial radii. Suppression of star formation (``quenching'') occurs in the infall regions of these structures even before galaxies enter the denser group environment.

\end{abstract}

 
 \section{INTRODUCTION}  
 \label{sec:INTRO}

The past three decades of observations and simulations have resulted in the paradigm of Lambda Cold Dark Matter ($\Lambda$CDM) hierarchical structure formation of the Universe, in which the clusters grow in size, mass and richness by accretion of isolated galaxies and galaxy groups along filaments (e.g., \citealp{1982ApJ...257..423H, 2006Natur.440.1137S, 2001MNRAS.328.1039C, 2000AJ....120.1579Y}). 

Simulations show that a significant fraction ($\sim$25$\%$-40$\%$) of galaxy accretion into halo masses ranging from $10^{12.9}$ to $10^{15.3} h^{\text{--}1}M_{\odot}$ between redshifts $z$ $\sim$0 to $z$ $\sim$1.5, occurs through groups rather than individual galaxies (e.g., \citealp{2009ApJ...690.1292B, 2009MNRAS.400..937M}).

Strong empirical evidence connects galaxy environment to galaxy properties, such as color, star-formation (SF) capabilities, morphologies, dynamics, etc. (e.g., \citealp{1999ApJS..122...51D}, \citealp{1951ApJ...113..413S,1997ApJS..110..213S, 2006ApJ...642..188P}), and that these properties vary significantly between galaxies in the `field'  and in the well-virialized clusters. The morphology-density relation of galaxies has proven to be remarkably robust both in clusters (up to redshift $z$ $\simeq$1) and in nearby groups (e.g., \citealp{1980ApJ...236..351D, 1997ApJ...490..577D, 1984ApJ...281...95P, 2005ApJ...623..721P, 2003ApJ...591...53T, 2005ApJ...620...78S}). Higher-density regions harbor red, older populations with significantly suppressed SF rates (SFR) compared to the lower density regions. The color-density relation appears at least as early as redshifts $z$ $\sim$1.5 (\citealp{2006astro.ph.12120C, 2007MNRAS.376.1445C, 2011A&A...527L..10F}). The SF-density relation and related galaxy properties, e.g., color-density, average stellar age-density, SF history-density, at least at low redshifts have been shown to exist not only within the cluster virial radius, but in the cluster outskirts, as well as in groups and the field (e.g., \citealp{1951ApJ...113..413S, 1998ApJ...499..589H, 2002MNRAS.334..673L, 2003ApJ...584..210G, 2004MNRAS.353..713K, 2004ApJ...615L.101B, 2002MNRAS.331..333P, 
2006MNRAS.373..469B, 2010MNRAS.404.1745M, 2011MNRAS.412.1098M}). In order to reconstruct galaxy transformation in the context of metallicity, color, morphology, SF history (SFH), AGN activity, etc., one needs a panchromatic approach to study galaxy transformations as a function of environment, and compare these data with the growing body of predictions from simulations.

\begin{figure}[h]
\includegraphics[scale=0.35, trim=0.55cm 3cm 1cm 6cm, clip=true]{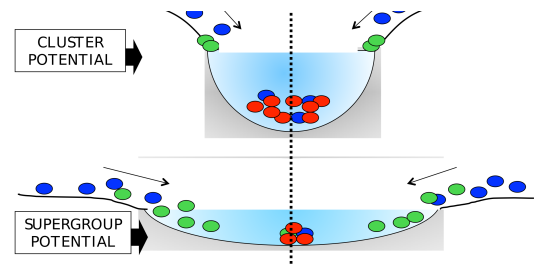} 
\caption{The schematic diagram shows galaxy evolution in a SuperCluster vs. a SuperGroup environment. A SuperGroup has a shallower but wider effective dark matter potential (bottom) compared to that of a Cluster (top). Blue and the red circles represent star-forming and non-star-forming galaxies, respectively. Green circles represent the transition state (`Green Valley') galaxies between the two. \label{fig:Galaxy-evol}} 
\end{figure}

Whether these transformations occur primarily as a result of a single dominant mechanism (e.g., ram pressure stripping due to the ICM), or as a combination of multiple mechanisms acting over various spatial- and temporal-scales is not clear. 
In relaxed clusters, several evolutionary mechanisms act on similar spatial- and temporal-scales, making it almost impossible to disentangle different local and global mechanisms, and ram pressure stripping appears to be the most dominant mechanism. An unrelaxed cluster or a cluster/filament precursor, on the other hand, has a shallower dark matter potential. Hence, the accreting galaxies are subjected to evolutionary mechanisms over larger  spatial- and temporal-scales, unlike in more virialized clusters ({\textit{Fig.~\ref{fig:Galaxy-evol}}}). This gives us a rare opportunity to study the early galaxy transformations that are otherwise difficult to disentangle once cluster-centric mechanisms begin to dominate. With extensive spectroscopy and photometric imaging of galaxies in the cluster outskirts, one can ideally map the radial locations of these transformations onto a time-sequence. This will help separate processes that are otherwise superimposed in rich cluster-filament interfaces, and hence, facilitate our understanding of interaction between filament and cluster-core at a different dynamical scale compared to a more evolved and relaxed system. \\

In this paper, we present a detailed photometric and spectroscopic map of the SuperGroup Complex of Abell 1882.

The questions we address in this work are stated as follows: (i) At which point during the early evolutionary history of the formation of a cluster does one see significant galaxy transformations that lead to the over-abundance of optically red galaxies that are observed at the core of the present-day clusters? In other words, are the well established color-color and color-density relations, seen in the present-day clusters, also seen in a unrelaxed cluster like Abell 1882? (ii) Is there evidence of galaxy transformation as a functions of both the number density of the galaxies and the spatial locations of the galaxies within the structure? (iii) Is the galaxy transformation dependent on the mass of the galaxy?
 
This paper is organized as follows. In Section \ref{subsec:SG1882}, we review results from the previous work on the SuperGroup environment of Abell 1882. In Section \ref{sec:OBS}, we describe the observations and data selection. In Section \ref{sec:SG1882}, we quantify the complex environment of Abell 1882. In Section \ref{sec:RESULTS}, we present the results and their implications. And in Section \ref{sec:SUM}, we present our main results.

The total current stellar masses and the k-corrections for this work have been obtained from the KCORRECT package \texttt{v3\_2} \citep{2003ApJ...594..186B}. 
The adopted cosmology for this work is $H_{0}$ = 71 km$s^{\text{--}1}$$Mpc^{\text{--}1}$, $\Omega_{M}$ = 0.27 and  $\Omega_{\Lambda}$ = 0.73 as recommended by CMB analysis \citep{2018arXiv180706209P}. All the positions are expressed in epoch J2000 coordinates.

 \subsection{\emph {SuperGroup Complex Abell 1882}} \label{subsec:SG1882}

 Even with only a handful of unrelaxed clusters observed at low redshifts (\citealp{2006MNRAS.370.1223B, 2003ApJ...584..210G, 2009ApJ...705..809T}), it is clear that these are diverse in structure. Hence, there may be several pathways to a virialized cluster, a `SuperGroup' being one such pathway. A SuperGroup is a ``group of groups'' of galaxies that are in the process of coalescing, and will eventually accrete enough mass to form a cluster. 
 
The SuperGroup Abell 1882 (\citealp{2003ApJS..146..267M, 2009POBeo..86..305E, 2013ApJ...772..104O}) is associated with a highly diverse filamentary large-scale environment. 
 
 It has a redshift of $z$ $=$ 0.139, and is centered approximately at $\alpha$ $=$ $14^{h}14^{m}39.9^{s}$ and $\delta$ $=$ \text{--}00$^{\circ}$19$'$57$''$ (J2000). It covers a much wider area than most previously observed intermediate-redshift relaxed clusters. Abell 1882 consists of at least three virialized groups, which form the core of the structure, and the filamentary outskirts indicate that it is still accreting mass. Hence, the local galaxy density and projected structure-centric distances are sufficiently decoupled to trace the galaxy evolution mechanisms in different environments within the system.

 This provides a unique low-redshift opportunity to explore the formation process of a cluster of galaxies from its original group/filament precursors. By comparing the kinematical data with N-body numerical simulations, the mass of the system is estimated to be  $M/M_{\odot}$$\sim$2$\times10^{14}$, which means that the system will coalesce into a Coma-like cluster in $\sim$2 Gyrs (\citealp{2010AAS...21535103G}); see also \citealp{2005ApJ...624L..73G, 2006MNRAS.370.1223B}.

 A combined optical and X-ray analysis by \citet{2013ApJ...772..104O} shows that two of the central galaxy groups appear to lie within 1\text{--}1.5 $r_{virial}$ (virial radius) from each other. This leads to a central dense region or the `swept-up' region (as defined by \citealp{Vijay}) between the three major groups, where the outer halos of the groups overlap. This `swept-up' region in Abell 1882 has higher galaxy number density than the outskirts, but more tenuous than the groups themselves. 

These factors strongly indicate that although Abell 1882 contains dynamically mature groups with their own individual well formed X-Ray structures, it is nonetheless a dynamically young system on larger scales.

 \section{OBSERVATION AND DATA REDUCTION}  
 \label{sec:OBS}
 

We have constructed the largest available optical and UV galaxy catalog for the Abell 1882 SuperGroup, containing 526 member galaxies based on their redshift distribution. This catalog contains spectra of galaxies from multiple galaxy groups that form the core of the SuperGroup, as well as the feeding filaments and the infall region. These contrasting environments provide us with an ideal laboratory for the study of galaxy evolution driven by structure formation ({\textit{Fig.~\ref{fig:radec2_d4000_nuv_ha}}}). We have complemented the optical spectroscopic data from MMTO/Hectospec at  Mt. Hopkins with data from the Sloan Digital Sky Survey (SDSS), the Galaxy And Mass Assembly (GAMA), the NASA/IPAC Extragalactic Database (NED) and Galaxy Evolution Explorer (GALEX). We also have data from targeted observation using the Gemini Multi-Object Spectrograph (GMOS) on Gemini South (\citealp{2010AAS...21641801M, 2010AAS...21535103G}). 
We then matched the final optical galaxy catalog with $NUV$ data from the Galaxy Evolution Explorer (GALEX) archive (GALEX program cycle: GI2 - 035, PI: Neal Miller). Table~\ref{tab:SURVEYS} lists the number of galaxies obtained from various surveys. Table~\ref{tab:CAT-SPEC} and Table~\ref{tab:CAT-PHOT} list the spectrometric and photometric data, respectively, for the galaxies in Abell 1882.\\

\begin{deluxetable*}{lcccL}[ht]
\tablecaption{Number of galaxies obtained from different surveys \label{tab:SURVEYS}}
\tablecolumns{5}
\tablenum{1}
\tablewidth{0pt}
\tablehead{
\colhead{Survey} &
\colhead{$\#$Galaxies With Spectroscopic Redshift} &
\colhead{Radius Of Sample} & \colhead{Radius Of Sample} & \colhead{AB Limit} \\
\colhead{\nodata} & \colhead{\nodata} &
\colhead{(degrees)} & \colhead{(Mpc)} & \colhead{(mag)}
}
\startdata
MMT - Hectospec &  210 &  0.54 & 4.74 & r $\leq$ 21 \\
Galaxy And Mass Assembly Survey & 170 &  1.05 &   9.14 &   r $\leq$ 19.8\\
Sloan Digital Sky Survey & 85 &  0.74 & 6.41 & r $\leq$ 17.8\\
Gemini Multi-Object Spectrographs & 38  &  0.17 & 1.5 & r $\leq$ 22\\
NASA/IPAC Extragalactic Database  & 23 &  0.68 & 6 & \nodata\\
GALEX(NUV) & 192 & 0.57& 4.97 & \nodata\\
\enddata

\end{deluxetable*}

\begin{figure*}[ht]
\plotone{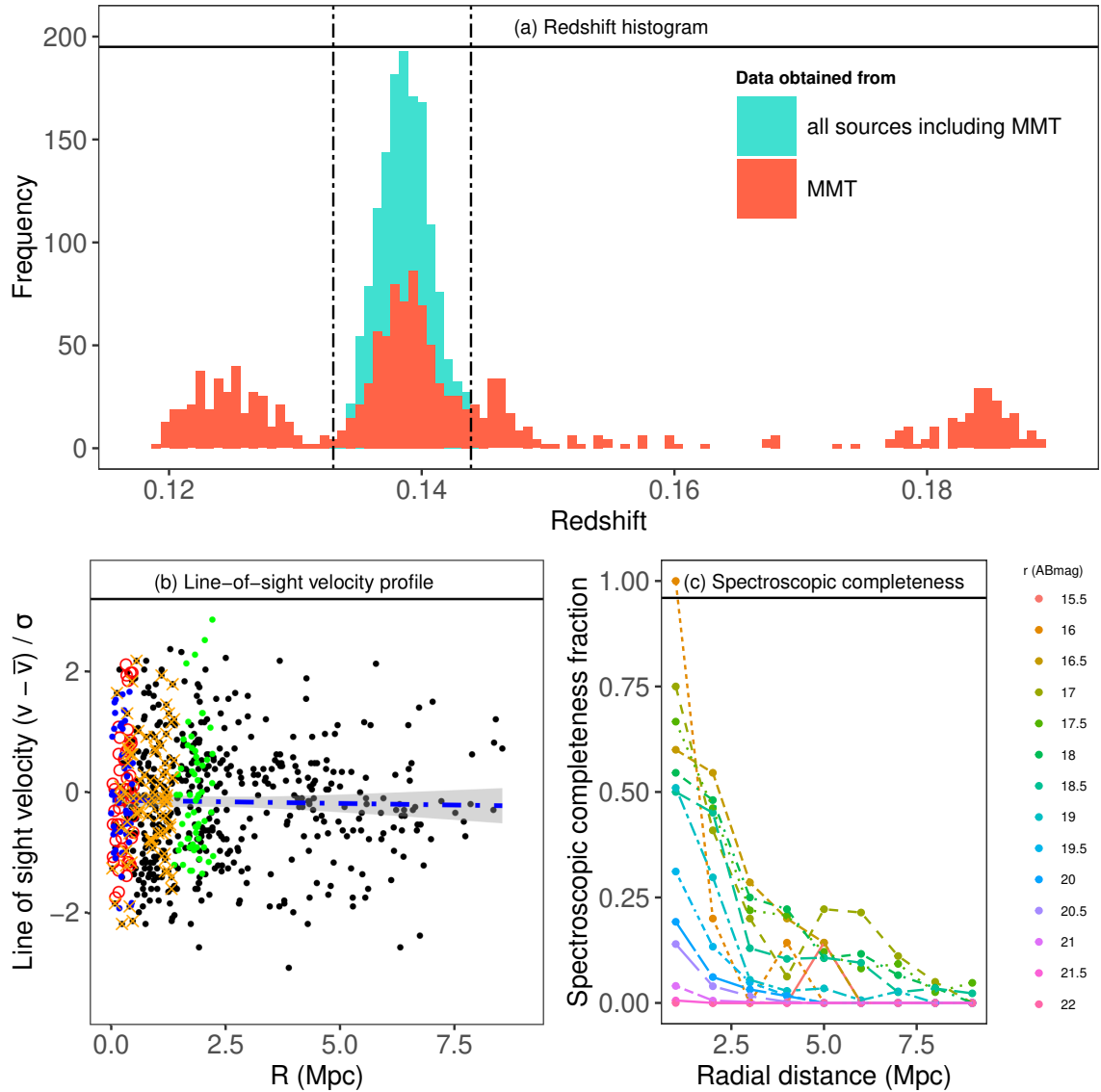}\caption{\textbf{\textit{(a)}} Redshift histogram of SuperGroup Abell 1882. \textbf{\textit{(b)}} Line-of-sight velocity profile of Abell 1882 galaxies. Red, blue and green circles represent the line-of-sight velocities for the galaxies in Group 1, Group 2 and Group 3, from their respective centers. Orange crosses represent the line-of-sight velocities for the swept-up region. The blue dashed line shows a linear model fit with $95\%$ CI. \textbf{\textit{(c)}} Spectroscopic completeness for Abell 1882 as a function of r-magnitude and the cluster-centric radius in Mpc.\label{fig:histo-caustic-cdf}}

\end{figure*}

\begin{deluxetable*}{lccccccc}[ht]
\tablecaption{Galaxy catalog of Abell 1882 (Spectral data) \label{tab:CAT-SPEC}}
\tablecolumns{8}
\tablenum{2}
\tablewidth{0pt}

\tablehead{
\colhead{Name} &
 \colhead{RA (J2000)} & 
\colhead{Dec (J2000)} & 
\colhead{Redshift} & 
\colhead{error(Redshift)} &
\colhead{EW(H$\alpha$)} &
\colhead{error(EW(H$\alpha$))}  & 
\colhead{$D_{n}4000$}
\\
\colhead{\nodata} &
\colhead{($^{\circ}$)}  &
\colhead{($^{\circ}$)} &
\colhead{\nodata}  &
\colhead{\nodata} &
\colhead{($\AA$)} &
\colhead{($\AA$)}  &
\colhead{\nodata} 
}
\startdata
1	&213.57009	&-0.40192		&0.138	&1.51E-05	&-11.81	&1.66		&1.38\\
2	&213.50627	&-0.22117		&0.136	&2.57E-05	&-20.49	&3.98	&1.13\\
3	&213.49960	&-0.33654		&0.136	&3.16E-05	&-11.81	&0.88	&1.87\\
4	&213.58050	&-0.44757		&0.139	&5.56E-05	&-38.64	&1.13		&1.16\\
5	&213.64855	&-0.36581		&0.140	&2.08E-05	&-21.21	&0.59	&1.18\\
6	&213.56176	&-0.42153		&0.135	&1.57E-05	&-0.646	&1.08		&1.77\\
7	&213.43641	&-0.41416		&0.139	&6.94E-05	&-26.34	&2.66		&1.31\\
8	&213.65039	&-0.33813		&0.138	&1.45E-05	&-10.53	&0.48	&1.45\\
9	&213.45533	&-0.42338		&0.138	&2.09E-05	&-4.292	&1.76		&1.47\\
10	&213.53138	&-0.39634		&0.140	&1.28E-05	&-1.544	&1.48	&1.60\\
\enddata

\end{deluxetable*}

\begin{deluxetable*}{lcccccl}[hbt!]
\tablecaption{Galaxy catalog of Abell 1882 (Photometric data) \label{tab:CAT-PHOT}}
\tablecolumns{7}
\tablenum{3}
\tablewidth{0pt}

\tablehead{
\colhead{Name} &
 \colhead{$u\text{--}r$} & 
\colhead{error($u\text{--}r$)} & 
\colhead{$NUV\text{--}r$} & 
\colhead{error($NUV\text{--}r$)} &
\colhead{Log(Mass)} &
\colhead{Source}  
\\
\colhead{\nodata} &
\colhead{(AB mag)}  &
\colhead{(AB mag)} &
\colhead{(AB mag)}  &
\colhead{(AB mag)} &
\colhead{\nodata} &
\colhead{\nodata}  
}
\startdata
1	&1.65	&0.01	&2.74	&0.08	&9.83	&MMT\\
2	&4.39	&0.09	&1.50	&0.14	&9.31	&MMT\\
3	&2.80	&0.02	&\nodata		&\nodata		&9.74	&MMT\\
4	&2.63	&0.02	&\nodata		&\nodata		&10.03	&MMT\\
5	&1.57	&0.05	&\nodata		&\nodata		&9.02	&MMT\\
6	&1.89	&0.02	&3.41	&0.18	&9.64	&MMT\\
7	&2.56	&0.10	&0.72	&0.16	&8.97	&MMT\\
8	&2.22	&0.01	&4.18	&0.13	&10.20	&MMT\\
9	&2.17	&0.01	&4.15	&0.23	&9.85	&MMT\\
10	&2.67	&0.00	&5.18	&0.11	&10.78	&MMT\\
\enddata

\tablecomments{Contact for the complete data set: \href{mailto:asengup@iu.edu}{asengup@iu.edu} }
\end{deluxetable*}

\subsection{MMTO/Hectospec} \label{subsec:MMTO}

The Hectospec fiber system on the 6.5m MMT (\citealp{2005PASP..117.1411F, 2008PASP..120.1222F}) can place 300 fibers over a field of about 1${^\circ}$, or 8.7 Mpc at the mean redshift of Abell 1882. This is large enough to include both the SuperGroup and the infall region, along with its feeding filaments. It samples a vast range of galaxy environments in a single MMT pointing. We targeted the galaxy population to $r$ $\leq$ 21 ($M_{\star}$ $+$ 4) and log($M_{\star}/M_{\Sun}$) $\geq$ 8.23. This enabled us to sample the faint magnitude, low-mass end of the galaxy mass spectrum out to a very large radius and deep into the feeding filaments, thus probing galaxy transformations in the far outskirts of the SuperGroup to very faint dwarf galaxies at this redshift. Selection priority for MMT observations used  24$\mu$m detections. Hence, FIR emitting galaxies in Abell 1882 have a higher probability of having MMT spectra than other galaxies in our sample set. We employed 5 fiber configurations with 30 sky-fibers, and 5 for spectrophotometric standards, leaving 265 target fibers per configuration. Given the possibility of probe collisions, a secondary target list was constructed in the same manner as the primary, but requiring weaker 24$\mu$m emission ($\sim$5$\sigma$). We used the 270 l/mm grating, yielding a spectral range of $3650-9200{\AA}$. One hour on-sky integration in three 20 minute exposure per configuration gave us SNR $>$ 5 for all emission line galaxies. Most of the 24$\mu$m sources have emission lines, making redshift estimates easier than from absorption spectral features. 
 
Galaxy redshifts were measured using Fourier Cross-Correlation (FXCOR) and the SPLOT task in IRAF\footnote{IRAF is distributed by the National Optical Astronomy Observatories, which are operated by the Association of Universities for Research in Astronomy, Inc., under cooperative agreement with the National Science Foundation.}.  We have applied FXCOR for the part of the spectrum shortward of the atmospheric oxygen line (5577{\AA}), which includes a sufficient number of  well-defined emission and absorption lines to provide an accurate galaxy redshift. Template spectra came from the Kennicutt atlas \citep{2004yCat.7141....0K}. Using the SPLOT task, we measured the wavelengths of significant absorption and emission lines: $H_{\alpha}$, $H_{\beta}$, [NII], CaII K $\&$ H, [OII] and [OIII] to manually determine the redshift of each galaxy individually. We then compared these redshifts with those obtained from the MMT pipeline. Values of redshifts obtained for the galaxies employing these different techniques are largely consistent.

 \subsection{Data from the Galaxy And Mass Assembly (GAMA) Survey} \label{subsec:GAMA}

The Galaxy And Mass Assembly (GAMA) survey \citep{2015MNRAS.452.2087L} builds on existing wide-field spectroscopic surveys like SDSS, 2dFGRS and the Millennium Galaxy Catalogue (MGC), down to r-magnitude $<$ 19.8 mag and $M^{\star}$$+$2 mag at $z$ $\sim$0.139. GAMA fiber diameter is 2.0" or 5.9 kpc at this redshift.

Emission and absorption lines were measured with SPLOT in batch mode after the lines are fitted using Gaussian profiles. The continuum for each line was selected using the recommendation in the SDSS archive. Only those lines which are within $\pm$$10{\AA}$ from the recommended line centers are used as detections in the redshift fitting. Several spectra from the GAMA survey have low S/N-ratio in the bluer wavelength regime, and hence, detection of H$\delta$ at $4100{\AA}$ and $[OII]$ at $3726{\AA}$ are not reliable for those spectra, and have been discarded. Manual measurements were done for equivalent width (EW) values that were obtained incorrectly by the IRAF script due to the discontinuity in the continuum of the spectra and/or low S/N-ratio. 

The spectra for the GAMA survey have been flux-scaled centered at $6300{\AA}$ ($6200{\AA}$ - $6400{\AA}$) to match that of the MMT for consistency.

 \subsection{Sloan Digital Sky Survey (SDSS)} \label{subsec:SDSS}

We used archival data from Data Release 12 (DR12) \citep{2015ApJS..219...12A}. We retrieved galaxy data centered at $\alpha$ $=$ $14^{h}14^{m}39.9^{s}$ and $\delta$ $=$ \text{--}00$^{\circ}$19$'$57$''$ (J2000) within a radius of 45$'$ and with a redshift range of 0.133 $<$ $z$ $<$ 0.144. 

The SDSS spectroscopic sample is incomplete for galaxies with nearby neighbors. This is due to the fact that the fibers cannot be very close together, typically less than 55$''$. Corrections for these missing objects are needed for the statistical correction of spectroscopic catalogs.

\subsection{GALEX Data} \label{subsec:GALEX}

The $NUV$ data were obtained from the GALEX archive (\citealp{2005ApJ...619L...1M, 2007ApJS..173..682M}) selected within 5$\arcsec$ radius of each optical galaxy. K-corrections were applied, and Galactic foreground extinction in the UV was applied using the recipe from \citep{1989ApJ...345..245C} using $A_{BV} = \left(a(\lambda) + \frac{b(\lambda)}{R_{v}}\right) \times E(B\text{--}V)$, where $R_{v}$ parametrizes the extinction law and $\lambda$ is the wavelength. For most locations in the Milky Way, $R_{v}$ = 3.1 is a representative value and the Galactic extinction is E(B\text{--}V) = 0.041 mag towards Abell 1882. We scaled to $\lambda$ = 2800{\AA} in the $NUV$, and found the Galactic extinction for the $NUV$ to be about 0.08 mag. We choose the $NUV$ for our analysis, because we have $\sim$36$\%$ of the $NUV$ data matching our optical catalog, whereas only $\sim$24$\%$ of the FUV data matches with our optical catalog. In addition, the $NUV\text{--}r$ color correlates more strongly with specific star-formation rate (sSFR) of galaxies than $FUV\text{--}r$ color (e.g., \citealp{2014SerAJ.189....1S}).\\

A reference cluster center position was determined by eye at roughly equal distances from the three major optical groups. The member galaxies are constrained within a redshift range of 0.133 $\leq$ $z$ $\leq$ 0.144 and a radius of 1.05${^\circ}$ or 9.14 Mpc from the adopted center of Abell 1882 at $\alpha$ $=$ $14^{h}14^{m}39.9^{s}$ and $\delta$ $=$ \text{--}00$^{\circ}$19$'$57$''$ (J2000), as shown in {\textit{Fig.~\ref{fig:histo-caustic-cdf}a}}. The red histogram represents all 1185 galaxy spectra obtained using the MMT Hectospec that have been used to constrain the  catalog. The blue histogram shows all additional member galaxies obtained from GAMA, the SDSS archive, NED and GMOS. Dashed vertical lines mark the adopted redshift cutoffs for Abell 1882 (0.133 $<$ $z$ $<$ 0.144). The galaxy groups are defined as circles with a diameter of $\sim$1 Mpc each with their centers in epoch J2000 coordinates at ($\alpha$ $=$ $14^{h}15^{m}07^{s}$, $\delta$ $=$ \text{--}00$^{\circ}$29$'$35$''$), ($\alpha$ $=$ $14^{h}14^{m}24^{s}$, $\delta$ $=$ \text{--}00$^{\circ}$22$'$46$''$) and ($\alpha$ $=$ $14^{h}14.07^{m}15^{s}$, $\delta$ $=$ \text{--}00$^{\circ}$21$'$00$''$), respectively, the centers for first two of which are adopted from \citet{2013ApJ...772..104O}, whereas, the third group was determined from the spatial distribution of optical galaxies in our catalog. A circle with a diameter of $\sim$3 Mpc encloses the three groups and the `swept-up' region. As expected in a virializing system, the infall and orbiting of galaxies together produce a near trumpet-shaped line-of-sight velocity profiles in the cluster region, as shown in {\textit{Fig.~\ref{fig:histo-caustic-cdf}b}}. The entire SuperGroup Abell 1882, with its groups and feeding filaments, has a velocity dispersion of 620 km s$^{-1}$. This value is lower than that for more massive and relaxed clusters like Coma cluster which have $\sigma$ $\simeq$ 1000 km s$^{-1}$ (e.g., \citealp{1970ApJ...162..333R}). The velocity dispersions of Group 1, Group 2 and Group 3, from their respective centers (not from the assumed center of Abell1881), are $\sim$669 km s$^{-1}$, $\sim$687 km s$^{-1}$ and $\sim$621 km s$^{-1}$, respectively (as shown in red, blue and green, respectively, in {\textit{Fig.~\ref{fig:histo-caustic-cdf}b}}). Orange crosses represent the line-of-sight velocities for the `swept-up' region. The blue dashed line shows a linear model fit with $95\%$ CI. The velocity dispersions of the galaxies in the filaments range from $\sim$490 km s$^{-1}$ for Filament 3, to $\sim$769 km s$^{-1}$ for Filament 2, and the velocity dispersion for Filament 1 is $\sim$550 km s$^{-1}$. The velocity dispersion of galaxies that are further away from the assumed center of the SuperGroup, and not a member of the galaxy groups, filaments or the `swept-up' region, is $\sim$543 km s$^{-1}$. The velocity dispersion of the galaxy groups is much higher than the galaxies in the lower density outskirts as expected. That is, the groups are significantly on the way to virialization, if not there.\\

\section{Quantifying Galaxy Environment Within Abell 1882} \label{sec:SG1882}

\subsection{Local Galaxy Density ($\Sigma$) Profile of Abell 1882} \label{subsec:LGD}

\begin{figure}[h] 

\includegraphics[scale=0.98, trim=0.cm 0cm 8cm 0.cm, clip=true]{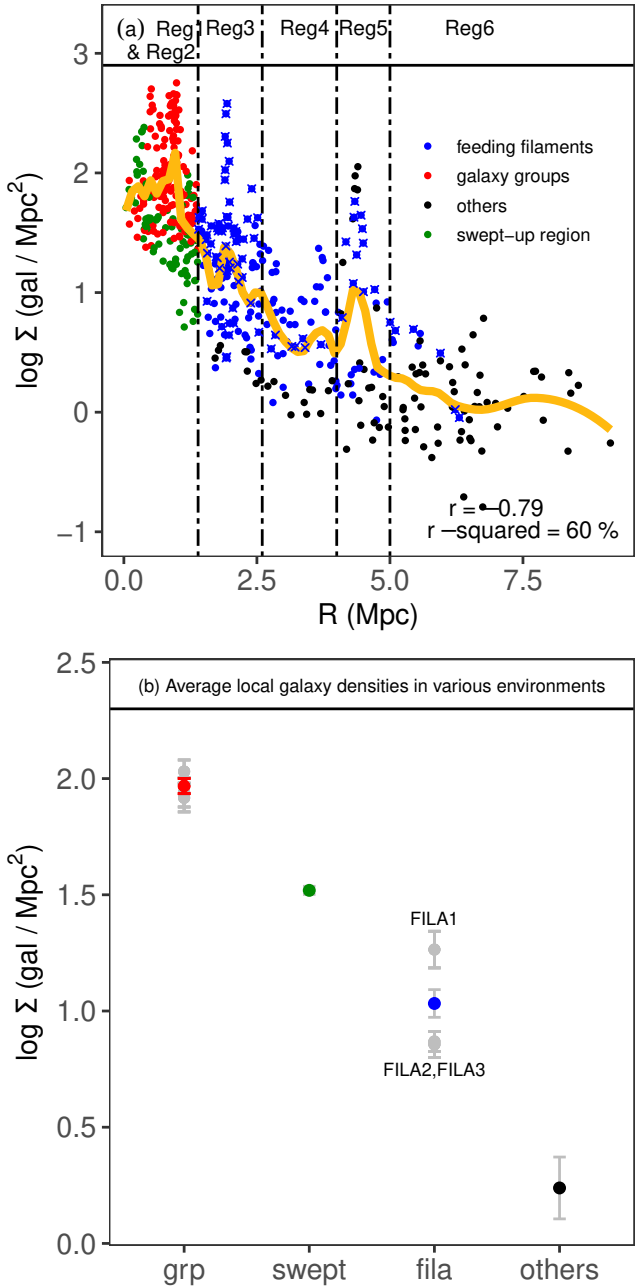} 
\caption{\textbf{\textit{(a)}} Correlation plot of the projected radial distance $\mathcal{R}$ from the adopted center of Abell 1882, and the local galaxy density $\Sigma$ of the galaxies. The yellow line represents the running mean of $\Sigma$. \textbf{\textit{(b)}} Average densities of galaxies in the three groups, `swept-up' region, feeding filaments, and the `others' (see Section \ref{subsec:LGD}, and Table~\ref{tab:SUBSTRUCTURE}), respectively, with standard error ($1\sigma$) bars. The average densities of galaxies within the three individual filaments are shown in grey.}
\label{fig:density_profile_plots}
\end{figure}

\begin{figure}[h] 
\includegraphics[scale=0.5, trim=0.cm 0cm 0 0, clip=true]{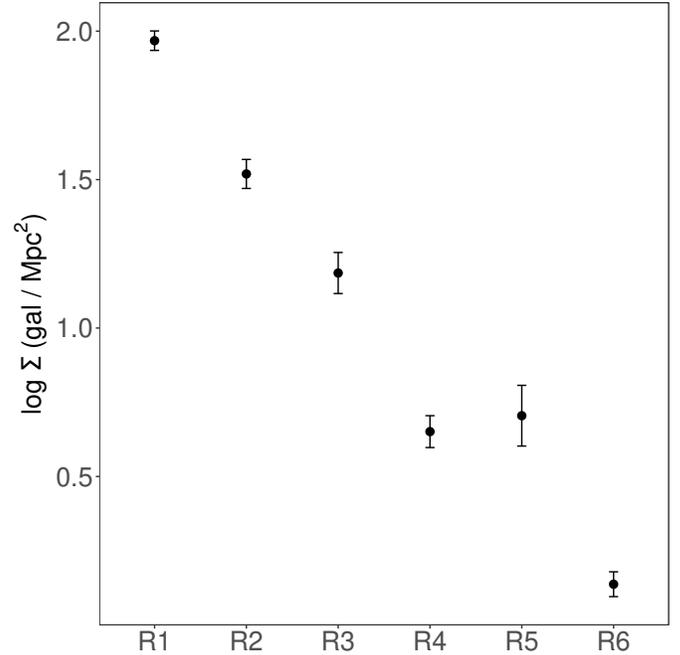} 
\caption{\textbf{\textit{(a)}} Correlation plot of regions (defined in Section \ref{subsec:ENV} and Table~\ref{tab:SUBSTRUCTURE2}), and $\Sigma$ in Abell 1882.}
\label{fig:avg_density_profile_byreg}
\end{figure}

We have calculated the projected local galaxy density (\lgd) in $Mpc^{\text{--}2}$ for each of the $N$ galaxies using the $n$-th nearest neighbor algorithm. In order to obtain a robust estimation of $\Sigma$, we follow \citet{2006MNRAS.373..469B}, where $\Sigma_{N}$ is given by $\frac{N}{\pi d_{N}^{2}}$, and $d_{N}$ is the averaged projected co-moving distance of a galaxy from its 4th and 5th nearest neighbors. 
 
 {\textit{Fig.~\ref{fig:density_profile_plots}a}} shows the trend of the projected local galaxy density $\Sigma$ with projected cluster-centric radius \rad using locally weighted scatterplot smoothing. The three major galaxy groups are shown in red, the galaxies in the `swept-up' region are shown in green, and the galaxies within the feeding filaments are shown in blue. 
Galaxies indicated in $black$ are member galaxies of Abell 1882 that do not lie within any of the galaxy groups, filaments or the `swept-up' region. We will refer to these galaxies as `others'. The yellow line represents the running mean of the $\Sigma$ at various projected radial distances. \\

In order to quantify the correlation between the local galaxy density ($\Sigma$) and the projected cluster-centric radius $R$, we first compute the $p$-value ($<$ $2.2 \times 10^{\text{--}16}$). This value is less than 0.05 or 5$\%$. Hence, this correlation is rather significant. We have further used the following guidelines for the magnitude of Pearson's product-moment correlation coefficient $r$ as defined in \citet{Evans}, where $r$  measures the strength of a linear correlation between two variables and its statistical significance as following:\\
\noindent \textbf{\textit{$\bullet$}} Positive and negative values of $r$ signify positive and negative correlations, respectively.\\
\noindent \textbf{\textit{$\bullet$}} 0.00$\text{--}$0.19 signifies $`very$ $weak'$ correlation\\
\noindent \textbf{\textit{$\bullet$}} 0.20$\text{--}$0.39 signifies $`weak'$ correlation\\
\noindent \textbf{\textit{$\bullet$}} 0.40$\text{--}$0.59 signifies $`moderate'$ correlation\\
\noindent \textbf{\textit{$\bullet$}} 0.60$\text{--}$0.79 signifies $`strong'$ correlation\\
\noindent \textbf{\textit{$\bullet$}} 0.80$\text{--}$1.00 signifies $`very$ $strong'$ correlation

The  correlation coefficient $r$ is $\text{--}$0.78, thus signifying a $`strong'$ negative correlation between log($\Sigma$) and $R$.\\

 \subsection{Identifying feeding filaments using Friends-Of-Friends Algorithm} \label{subsec:FOF}

\begin{figure*}
\centering 
\includegraphics[scale=1]{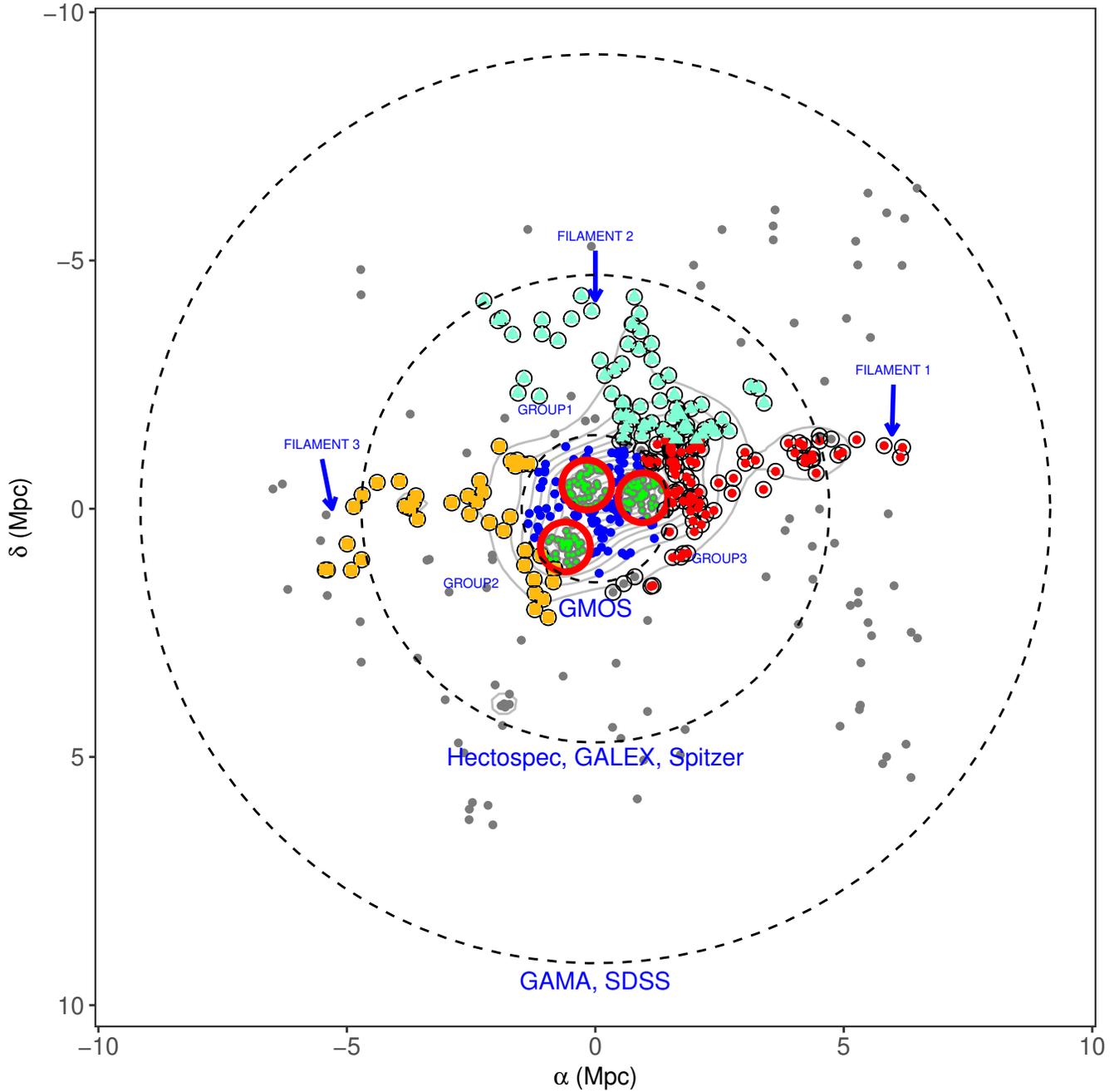} 
\caption{The three major filaments detected by the Friends-Of-Friends algorithm are shown in red dots, aquamarine triangles and yellow squares, and also indicated by blue arrows. Green dots encircled in red represent the galaxy group members. Blue dots represent the `swept-up' region. Grey dots represent all the galaxies categorized as `others'. The contours show a galaxy number density map. The dotted circles represent the area covered by various surveys used in the current work.} 
\label{fig:radec2.eps}
\end{figure*}

\begin{figure*}
\centering 
\includegraphics[scale=1]{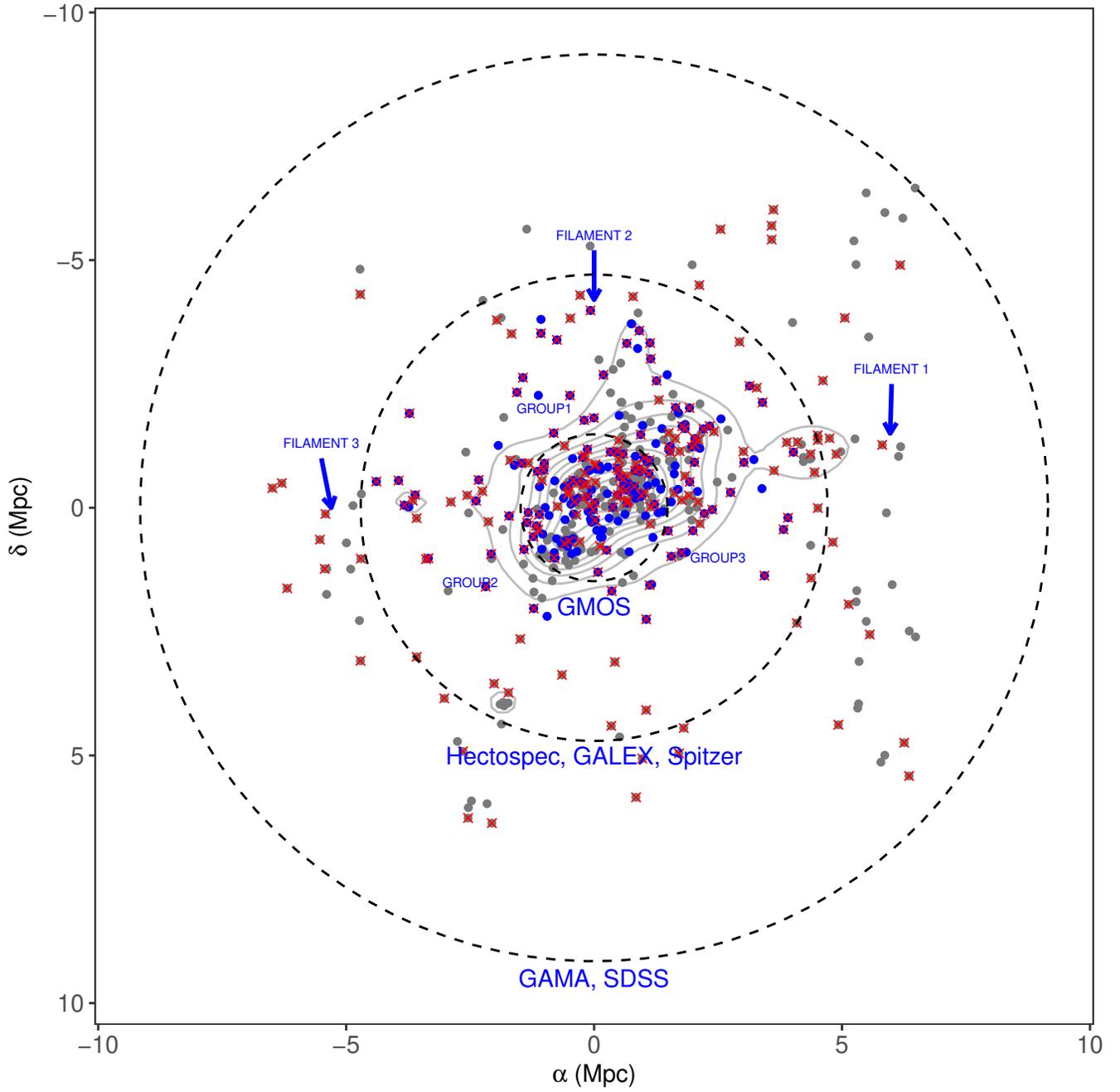} 
\caption{The optical galaxies overlaid with galaxies detected in $NUV$ (blue dots), and galaxies with $H_{\alpha}$ data (in red crosses). The contours show a galaxy number density map.}
\label{fig:radec2_d4000_nuv_ha}
\end{figure*}


\begin{figure}[h]
\centering 
\includegraphics[scale=0.6, trim=2cm 1cm 1cm 1cm, clip=true]{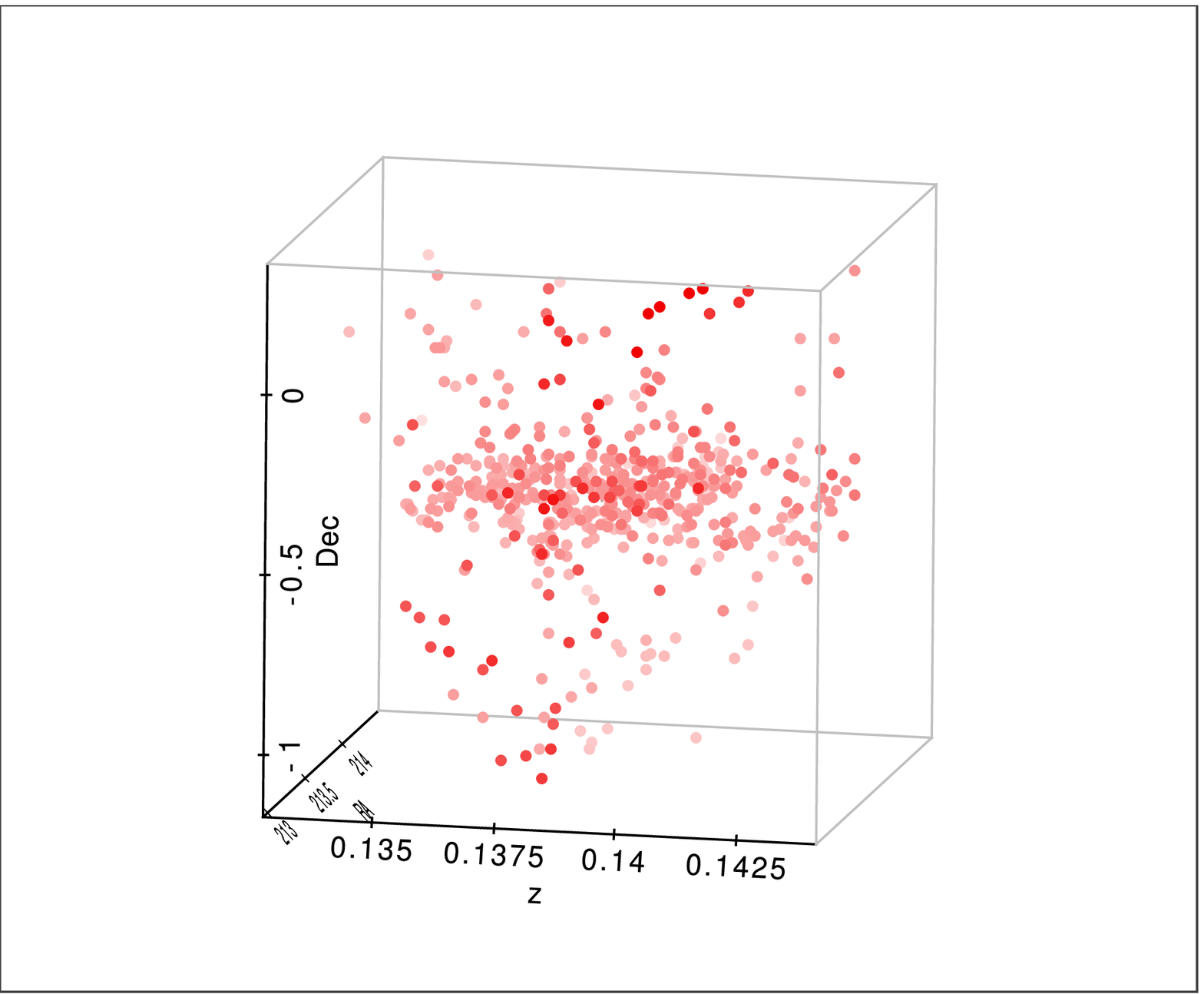} 
\caption{A redshift-declination slice of SuperGroup Abell 1882 with redshift. The dimmer red circles represent galaxies that are further away from that vantage point.
\label{fig:ra_dec_z}}
\end{figure}

We have used group searching code $\texttt{groups\_pro.c}$ (Ivan Valtchanov, 1998) based on the Friends-Of-Friends (FOF) percolation algorithm developed by \citet{1982ApJ...257..423H} to identify the feeding filaments in the cluster outskirts ({\textit{Fig.~\ref{fig:radec2.eps}}}). This method recursively links all galaxies satisfying the linkage based on distance between the galaxies and their velocities. 
It then finds galaxy number density enhancements for galaxies with spectroscopic redshifts, thus identifying the galaxies that form the filaments. We define a filament as a structure containing at least three galaxies connected by the linkage parameters. We have detected three feeding filaments that are shown in red circles, aquamarine triangles and yellow squares, indicated by blue arrows in {\textit{Fig.~\ref{fig:radec2.eps}}}. See also Table~\ref{tab:SUBSTRUCTURE} for more details. 
{\textit{Fig.~\ref{fig:radec2_d4000_nuv_ha}}} shows the optical galaxies (grey dots), overlaid with galaxies detected in $NUV$ (blue dots), and galaxies with $H_{\alpha}$ data (in red), in our catalog. Most of the $NUV$ galaxies in the catalog lie within the groups, or in their immediate infall region. 

Tying this result back to Section \ref{subsec:LGD}, {\textit{Fig.~\ref{fig:density_profile_plots}b}} shows the mean $\Sigma$ within each of the three groups, the `swept-up' region, the feeding filaments, and the `others'. As expected, the groups have the highest local galaxy densities and have comparable values. The `swept-up' region exhibits a lower $\Sigma$ than the groups, but is higher than the average $\Sigma$ within the filaments. The galaxies that lie outside these structures, i.e., the `others', have the lowest $\Sigma$, as expected. 

As mentioned in Section \ref{subsec:LGD}, we observe several over-densities as we move outwards from the central galaxy groups. Most of these over-densities lie along Filament 1. This is clear from {\textit{Fig.~\ref{fig:density_profile_plots}b}}, where the average local galaxy densities of the individual filaments are shown in grey. The average $\Sigma$ of Filament 1 is significantly higher than the other two filaments, as can also be seen in {\textit{Fig.~\ref{fig:density_profile_plots}a}} (blue crosses). The density contour (in grey) in {\textit{Fig.~\ref{fig:radec2.eps}}} points at a possibility of an yet undetected smaller galaxy group, or a galaxy group that is in the early stages of assembly within Filament 1, further to the right of the central galaxy groups. 

{\textit{Fig.~\ref{fig:ra_dec_z}}} shows a sideways view of SuperGroup Abell 1882 with redshift. It can be clearly seen that the galaxies are being accreted asymmetrically, mostly along the direction of the Right Ascension, which agrees well with the position of the majority of the detected feeding filaments. \\

\begin{table*}[ht]
\renewcommand\thetable{4} 
\caption{Number of galaxies, local galaxy densities, and velocity dispersion for each sub-structure within Abell 1882 \label{tab:SUBSTRUCTURE}}
\centering
\begin{tabular}{p{0.2\linewidth}p{0.15 \linewidth}p{0.15\linewidth}p{0.15\linewidth}p{0.15\linewidth}}
 \hline
\hline
Structure description & No. of galaxies & No. of galaxies  & $\Sigma$    & Velocity dispersion \\
 \nodata &  (in optical)  & (in $NUV$) & $Mpc^{2}$)  &  (km s$^{-1}$)\\ 

\hline
\hline
All environments in Abell 1882 & 526     &  191  &55 & 620 \\
\hline
All galaxy groups &150 &  58   & 133 & 664 \\
- Group 1  &  49 & \nodata&112  & 669\\
- Group 2 & 43 & \nodata &134 & 687\\
- Group 3 & 58 &\nodata &150 & 621\\
\hline
`Swept-up' region & 84&42&48 & 588\\
\hline
All three filaments & 194 & 78&22 & 634\\
- Filament 1  & 80 &  \nodata & 38 & 551\\
- Filament 2 & 66 & \nodata &11 & 769\\
- Filament 3  &  41 & \nodata  &10 & 490\\
\hline
`Others' & 97& 13&6 & 543\\
\hline
\end{tabular}
\end{table*}

\subsection{Total Current Stellar Masses} \label{subsec:MASS}

Total current stellar mass has been obtained from Bell and de Jong stellar mass model \citep{2001ApJ...550..212B} using the IRAF KCORRECT package, which uses the SDSS absolute magnitudes. Photometric zero point corrections, Galactic extinction corrections and k-corrections have been applied to the observed magnitudes. We have divided the galaxies into three mass bins: $high$ $mass$ galaxies (M $\ge$ $10^{10.5}$$M_{\odot}$), $intermediate$ $mass$ galaxies ($10^{9.5}$$M_{\odot}$ $\leq$ M $\leq$ $10^{10.5}$$M_{\odot}$), and the $dwarf$ $galaxies$ (M $<$ $10^{9.5}$$M_{\odot}$). The motivation for these mass divisions is that several studies have drawn attention to galaxies in the mass range of M $< 10^{9.5}$$M_{\odot}$, which show especially strong environmental transformations. The dwarf galaxies have been defined by the recipe provided by \citet{2001AJ....121.2358B}, or $z_{AB}$ $\ge$ 15 mag and $M_{z}$ $>$ $M^{\ast}$ + 2.3 mag for Coma Supercluster, where $M_{z}^{\ast}$ = \text{--} 22.32 mag. Using the above recipe, we get $z_{AB}$ $>$ 19.1 mag for dwarf galaxies at $z$=0.139. The low mass galaxies correspond reasonably well with this $z$-mag limit defined for dwarf galaxies in the Coma cluster. 

The MMT sampling goes fainter than the SDSS and GAMA survey limits, but covers only $1^{\circ}$ in the sky. We have sampled further out to $\sim$$2^{\circ}$ in the sky using the SDSS and GAMA surveys. Hence, the dwarf galaxy sampling is sparser in the outskirts.

We further divide the galaxy environments into three different regions as indicated in {\textit{Fig.~\ref{fig:density_profile_plots}}(a)} in black vertical dot-dashed lines. The regions are defined in \ref{subsec:ENV}.\\

\section{Results And Discussions} \label{sec:RESULTS}

\subsection{Defining The Different Environments For Comparison Of Galaxy Evolution Indicators} \label{subsec:ENV}

\begin{table*}[ht]
\renewcommand\thetable{5} 
\caption{Number of galaxies, local galaxy densities, and velocity dispersion for each region within Abell 1882 as defined in Section \ref{subsec:ENV} \label{tab:SUBSTRUCTURE2}}
\centering
\begin{tabular}{p{0.2\linewidth}p{0.15 \linewidth}p{0.15\linewidth}p{0.15\linewidth}p{0.15\linewidth}}
 \hline
\hline
Region & Radial distance   & Number of galaxies detected in optical  & Number of galaxies detected in $NUV$  & Local galaxy density $(\Sigma)$ \\
 \nodata &   (in Mpc) & \nodata  &  \nodata  &  (galaxies/$Mpc^{2}$)\\
\hline
\hline
Region 1 & $\mathcal{R}$ $< $1.4    &  150  &58 & $\sim$133\\
\hline
Region 2 &  $\mathcal{R}$ $< $1.4   & 84  & 42& $\sim$48\\
\hline
Region 3  &  1.4 $<$ $\mathcal{R}$ $<$ 2.6  &121 &53 & $\sim$29\\
\hline
Region 4 & 2.6  $<$ $\mathcal{R}$ $<$ 4  & 58& 34 & $\sim$7\\
\hline
Region 5 & 4  $<$ $\mathcal{R}$ $<$ 5  & 47 &4 & $\sim$17\\
\hline
Region 6 & $\mathcal{R}$ $>$ 5  & 65 & \nodata & $\sim$2\\
\hline
\end{tabular}
\end{table*}

Our catalog for Abell 1882 covers a vast range within its density and velocity fields, from the dense central core to the sparse distant outskirts from where the galaxies pour into the central potential of the SuperGroup along the filaments. This scenario provides us with four distinct galaxy environments as discussed in the previous sections, and as shown in {\textit{Fig.~\ref{fig:radec2.eps}}}: (i) the group environments (ii) central dense region, or the `swept-up' region between the three major groups, where the outer gaseous halos of the groups overlap, (iii) the feeding filaments as inferred from the FOF percolation algorithm, and (iv) the infalling galaxies that are not within the filament-environment (categorized as `others'). 

From Table~\ref{tab:SUBSTRUCTURE}, $\Sigma$ in the galaxy population categorized as `others' is the lowest within the environment of Abell 1882 with a density of about 6 galaxies/$Mpc^{2}$. This population also exhibits a much lower velocity dispersion compared to the galaxy groups. The average $\Sigma$ in the three filaments and the `swept-up' region are 22 galaxies/$Mpc^{2}$ and 48 galaxies/$Mpc^{2}$, respectively. Galaxy groups have the highest $\Sigma$, as expected (133 galaxies/$Mpc^{2}$). Filament 1, which shows evidence of over-dense regions in the far outskirts, has a much higher $\Sigma$ compared to the other filaments. This filament is connected to Group 3, which also exhibits the highest $\Sigma$ of all the three galaxy groups. This reinforces the conclusion that the accretion of the galaxies into Abell 1882 is highly asymmetric. Filament 1 also shows a much higher velocity dispersion compared to the other filaments, and also the galaxy groups, indicating an yet undetected smaller galaxy group, or a galaxy group that is in the early stages of assembly, as mentioned in the previous section.

However, we observe distinct over-densities at various \rad values all the way out to about 7 Mpc, indicating that this SuperGroup is highly clumpy, which strongly suggests that it is still in the process of growing and accreting today. 
This clumpiness could potentially help in disentangling the roles of $\Sigma$ and the radial positions of the galaxies in this SuperGroup environment. Hence, in addition to comparing the galaxy properties in the structures mentioned above, we will also trace the galaxy properties as a function of the radial distance from the assumed center of Abell 1882 as shown with black dashed vertical lines in {\textit{Fig.~\ref{fig:density_profile_plots}a}} (also see Table~\ref{tab:SUBSTRUCTURE2}). This classification of the regions have been carefully chosen in order to study the effect of the projected radial distances of the galaxies on their properties, as well as the effect of $\Sigma$, and the underlying large scale structures. Note that the galaxies in the filaments have been sampled in a manner so that they capture the over-densities in the filaments at various radial distances from the assumed center.\\
\noindent \textbf{\textit{$\bullet$}} Region 1 and Region 2 coincide with the groups and the `swept-up' regions, respectively, both of which lie within a radius of 1.4 Mpc from the assumed center of the SuperGroup.\\
\noindent \textbf{\textit{$\bullet$}} Region 3$\textit{--}$6 contain galaxies within the filaments as well as galaxies that are outside it (`others'). \\
\noindent \textbf{\textit{$\bullet$}} Region 3 lies between the radii of 1.4 to 2.6 Mpc. This region mostly consists of galaxies in the filaments as shown in blue in {\textit{Fig.~\ref{fig:density_profile_plots}a}}. There is a distinct over-density of galaxies in this region.\\
\noindent \textbf{\textit{$\bullet$}} Region 4 lies between the radii of 2.6 to 4 Mpc. This region  has a significantly lower $\Sigma$ than Region 3 as shown in {\textit{Fig.~\ref{fig:avg_density_profile_byreg}}}. \\
\noindent \textbf{\textit{$\bullet$}} Region 5 lies between the radii of 4 to 5 Mpc. This region shows an over-density in the filaments at this radial distance from the assumed center of Abell 1882, and has a higher $\Sigma$ than Region 4.\\
\noindent \textbf{\textit{$\bullet$}} Region 6 lies beyond a radius of 5 Mpc, and has the lowest $\Sigma$. It is dominated by galaxies categorized as `others'. \\

We will examine various parameters like $u\text{--}r$ $\&$ $NUV\text{--}r$ colors, and EW[$H\alpha$] with various galaxy environments described in this section. We will then narrow down the possible locations for quenching of galaxies that lead to the bimodality of colors that we see in the Abell 1882 SuperGroup. In the following sections, the NUV\text{--}r color evolution inferences for the red galaxies are drawn only from the $NUV$ bright galaxies, because of the incompleteness of the sample at the fainter magnitudes. Our sample in the $NUV$ is restricted within a radius of 4.7 Mpc from the adopted center. Hence, the $NUV$ faint galaxies and/or very far outskirts are poorly represented.

\subsection{Optical, UV Color-Magnitude \& Color-Mass Relations In Abell 1882} \label{subsec:CMD}

\begin{figure*}
\centering 
\includegraphics[scale=1, trim={0 9.3cm 0 0}, clip=true]{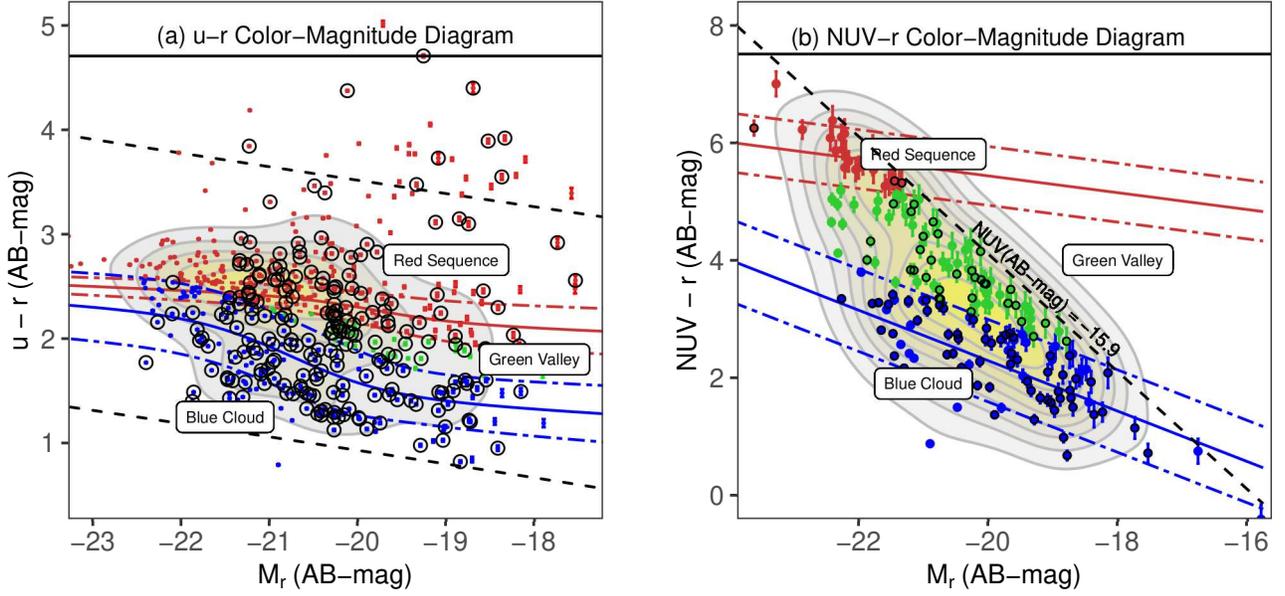} 
\caption{\textbf{\textit{(a)}} $u\text{--}r$ color-Magnitude Relation and  \textbf{\textit{(b)}} $NUV\text{--}r$ color-Magnitude Relation in Abell 1882. The black dashed lines represent the $95\%$ prediction interval (PI) for all galaxies in the sample set. The red and blue lines represent best fits for the Red Sequence and the Blue Cloud. Grey dots represent the galaxies in Abell 1882 in the color-color diagram. The black circles are the $H_{\alpha}$ emitters. The gray dashed line in fig(b) shows the NUV magnitude limit (= -15.9) for Abell 1882.}
\label{fig:cmd}
\end{figure*}

 \begin{figure*}
\centering 
\includegraphics[scale=1, trim={0 8.7cm 0 0}, clip=true]{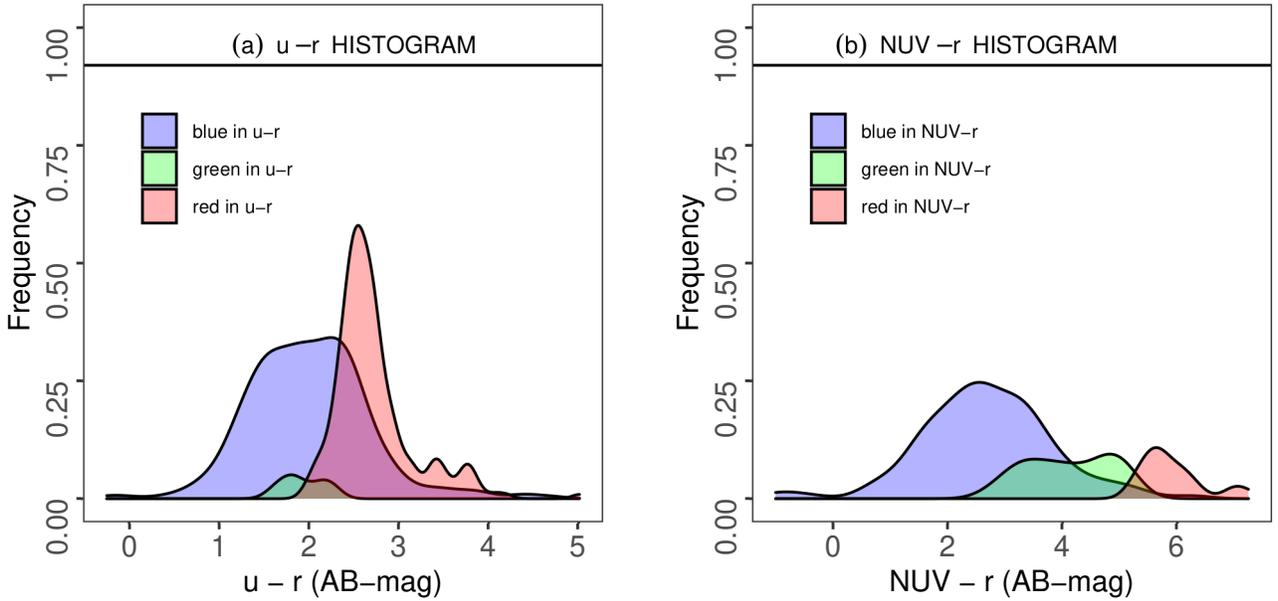} 
\caption{Histograms for red, blue and green galaxies, shown in red, blue and green colors, respectively, in \textbf{\textit{(a)}} $u\text{--}r$ and \textbf{\textit{(b)}} $NUV\text{--}r$ colors.} 
\label{fig:hist_with_color2}
\end{figure*}

\begin{figure*}
\centering 
\includegraphics[scale=1, trim={0 0 0 0}, clip=true]{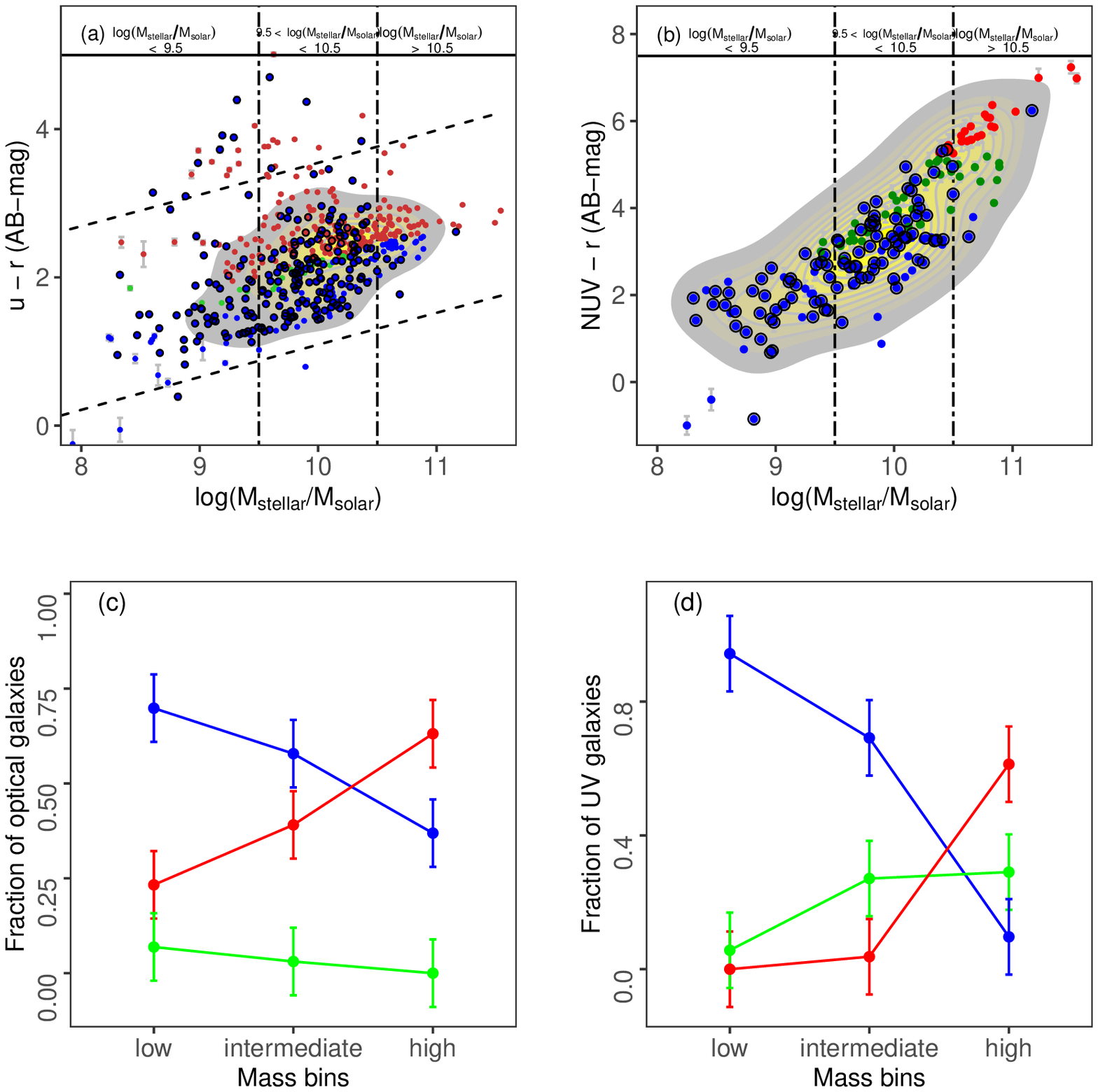} 
\caption{\textbf{\textit{(a)}} $u\text{--}r$ color-Mass Relation and \textbf{\textit{(b)}} $NUV\text{--}r$ color-Mass Relation in Abell 1882. Red, blue, and green dots represent the Red Sequence, Blue Cloud, and Green Valley galaxies, respectively. Black open circles represent the $H_{\alpha}$ emitters, all of which have been reclassified as blue galaxies. The black dashed lines represent the $95\%$ prediction interval (PI) for all galaxies in the sample. \textbf{\textit{(c)}} Fraction of optical galaxies, and \textbf{\textit{(d)}} fraction of $UV$ galaxies for different mass ranges. Galaxy mass bins are defined in Section \ref{subsec:MASS}. The Red Sequence, Blue cloud and Green Valley galaxies are represented by red, blue and green filled circles, respectively, using the new classification based on optical and UV colors, as well as $H_{\alpha}$ emissions.} \label{fig:colormass}
\end{figure*}

In {\textit{Fig.~\ref{fig:cmd}a\textit{--}b}}, the best fits for Red Sequence (RS) galaxies (red lines) and Blue Cloud (BC) galaxies (blue lines) have been plotted in the $u\text{--}r$ using the prescription from \citet{2004ApJ...600..681B} and \citet{2004ApJ...615L.101B}. For the $NUV$ galaxies, the best fits ($\textit{--}$ 0.143$M_{r}$ + 2.57) $\pm$0.5 and ($\textit{--}$ 0.429$M_{r}$$\textit{--}$ 6.29) $\pm$0.7 have been used for identifying red and blue galaxies, respectively, using the prescription from \citet{Yi2005}.

The flux-to-mass transformation is increasingly affected by details of the exact SFH as we consider shorter SF time scales. The $NUV$ arises from the photospheres of O-through late-type B-stars and early type A-stars with mass M $\geq$ $3M_{\odot}$. Hence, $NUV$ measures the SF averaged over the lifetimes of these stars i.e., past Gyr \citep{2007ApJS..173..619K}. This makes $NUV$ far more sensitive to recent and on-going SF compared to the optical colors of the galaxies, which traces SF over a period of $\sim$2$\text{--}$3 Gyrs. The $H_{\alpha}$ emitters are O-stars and early type B-stars with masses in excess of 17$M_{\odot}$. These stars have a lifetime of a few million years. Hence, the $H_{\alpha}$ emitting galaxies have undergone more recent significant SF episodes in the past $10^{7}$ years. $H_{\alpha}$ is sensitive to only high level of continuous SF. Whereas, the galaxies showing $NUV$ emission, but no $H_{\alpha}$ emission, have recent, low-level SF.

We have re-classified the red, blue, and green galaxies based on their optical and UV colors, as well as their EW[$H_{\alpha}$]. The red and green galaxies in optical and UV colors, with EW[$H_{\alpha}$] $<$ $\text{--}2{\AA}$, are possibly dusty starburst galaxies, and hence, have been re-classified as blue galaxies in both the optical and UV colors. We have used this new classification for all the figures in this paper, except in the color-magnitude diagrams ({\textit{Fig.~\ref{fig:cmd}a\text{--}b}}).\\

Abell 1882 has a well-defined optical Red Sequence (RS), Green Valley (GV), and Blue Cloud (BC) galaxies ({\textit{Fig.~\ref{fig:cmd}a}}). Although most of the galaxies fall within the $95\%$ Prediction Interval (PI) for all galaxies in the Abell 1882 catalog (black dashed lines), we see some RS galaxies that lie outside the PI especially towards the reddest region of the $u\text{--}r$ color, all of which are low to intermediate mass galaxies with $log$ $M/M_{\odot}$ $<$ $10.5$ ({\textit{Fig.~\ref{fig:colormass}a}}). Some of these red galaxies are also $H_{\alpha}$ emitters and have been reclassified as blue galaxies (shown with black circles in {\textit{Fig.~\ref{fig:colormass}a}}), thus signifying dusty starburst galaxies that have undergone recent SF in past few million years.  About $90\%$ of the galaxies in the red $u\text{--}r$ histogram in {\textit{Fig.~\ref{fig:hist_with_color2}a}} have mass $log$ $M/M_{\odot}$ $>$ $10.5$. 

Abell 1882 also exhibits a clear bimodality in the $NUV$-optical color-magnitude diagram, as shown in {\textit{Fig.~\ref{fig:cmd}b}}. 

We classify the galaxies in Abell 1882 into the three following sequences based on the $NUV\text{--}r$ color-magnitude diagram:

(i) Passive or the RS, i.e., galaxies that are red in $NUV\text{--}r$ color (red filled circles): The low luminosity galaxies in $NUV,$ especially those on the RS, are significantly undercounted due to the GALEX flux detection threshold. Hence, by selection, all the red galaxies on the RS represent only the massive, $NUV$ bright galaxies ($log$ $M/M_{\odot}$$>$ 10.4). In this sequence, only $7.5\%$ of the galaxies are $H_{\alpha}$ emitters, indicating that most of these galaxies had no recent star-forming episodes in past 1 Gyr. These recent star-forming galaxies have been re-classified as star-forming sequence or the BC (blue filled circles). Henceforth, RS, or red galaxies will indicate only the $u\text{--}r$ or $NUV\text{--}r$ red galaxies that are not $H_{\alpha}$ emitters.

(ii) Star-forming sequence or the BC (blue filled circles): These galaxies are mostly blue in their $NUV\text{--}r$ color, indicating star-forming episodes in at least past 1 Gyr. About $67\%$ of the galaxies in this sequence are $H_{\alpha}$ emitters, and hence have undergone recent significant SF in past $10^{7}$ years. Henceforth, Blue Cloud, or blue galaxies will indicate the $u\text{--}r$ or $NUV\text{--}r$ blue galaxies, in addition to the $H_{\alpha}$ emitting galaxies that were photometrically categorized as red or green.

(iii) The intermediate phase or the GV galaxies (green filled circles): This phase represents a transitional phase between the Passive sequence and the Star-forming sequence (See   \citealp{Salim07, 2005MNRAS.360...60K, 2007ApJS..173..293W, 2007ApJS..173..342M,2014MNRAS.440..889S}). These galaxies have undergone recent low-level SF. However, a number of the GV galaxies in $NUV\text{--}r$ color show $H_{\alpha}$ emission (black open circles in {\textit{Fig.~\ref{fig:cmd}b}}). These galaxies are possibly dusty starburst galaxies, indicating only recent, ongoing starburst episodes (in the past few million years), and hence undetected in the $NUV$ color. These recent, star-forming galaxies have been re-classified as star-forming sequence or the BC (blue filled circles). Henceforth, GV, or green galaxies will indicate only the $u\text{--}r$ or $NUV\text{--}r$ green galaxies that are not $H_{\alpha}$ emitters.

There are several pathways for galaxies to transition through the GV: (a) The galaxies could be transitioning from the BC to the RS as they are quenched; (b) Minor, gas-rich mergers in passive galaxies \citep{2005MNRAS.360...60K}; (c) RS galaxies may accrete gas from the intergalactic medium and undergo low-level SF \citep{2007ApJS..173..572T}.

{\textit{Fig.~\ref{fig:colormass}c\text{--}d}} show the fractional numbers of optical and $NUV$ galaxies for the three different mass ranges. The high mass, intermediate-mass, and the dwarf galaxies are defined in Section \ref{subsec:MASS}. The RS, BC and GV galaxies, as described above, are represented by red, blue and green filled circles, respectively.

In both the optical and UV color, the high mass population ($log$ $M/M_{\odot}$ $>$ $10.5$) is dominated by the RS galaxies ($\sim$63$\%$ and $\sim$61$\%$, respectively). However, in stark contrast to the optical colors, the $NUV$ color reveals a large number of GV galaxies (41 non-$H\alpha$ emitting GV galaxies in $NUV$, compared to only 18 in optical color), most of which have $log$ $M/M_{\odot}$ $>$ $9.5$. These are low-level recent star-forming galaxies in the past Gyr. This leads to a significantly more pronounced GV in $NUV\text{--}r$ color, in spite of undercounting the faintest galaxies due to the GALEX magnitude limit. Hence, $NUV$ is more effective in detecting low-level SF compared to its optical counterpart, and has revealed a significant number of low-level star-forming galaxies that were undetected in the optical color.  Only $\sim$2$\%$ of the $NUV$ blue galaxies are high mass galaxies ($log$ $M/M_{\odot}$ $>$ $10.5$). This indicates that most of the massive galaxies have already moved to the RS in past Gyr, although  there is also a significant number of high mass $NUV$ GV galaxies, indicating low-level SF in the massive galaxies in the past Gyr.

In the intermediate mass range, there is a larger fraction of blue galaxies in both the optical and UV colors. In addition, a large number of green intermediate mass galaxies have been revealed. About one-third of the high mass galaxies and intermediate mass galaxies detected in $NUV$ are GV galaxies, and hence have undergone a change in their star-forming pattern (i.e., low-level SF episodes) in the past Gyr. The $u\text{--}r$ color does not reflect these recent low-level SF episodes in the massive galaxies.

The dwarf galaxies have significantly more optically blue than red galaxies. The fractional numbers of the dwarf galaxies in $NUV\text{--}r$ color are affected by the GALEX flux threshold, which severely undercounts the red and green dwarf galaxies.

\subsection{Spatial Distribution Of Star-forming, Passive, $\&$ Transitioning Galaxies}\label{subsec:SPATIAL}

\begin{figure*}
\centering 
\includegraphics[scale=1, trim={0 0 0 0}, clip=true]{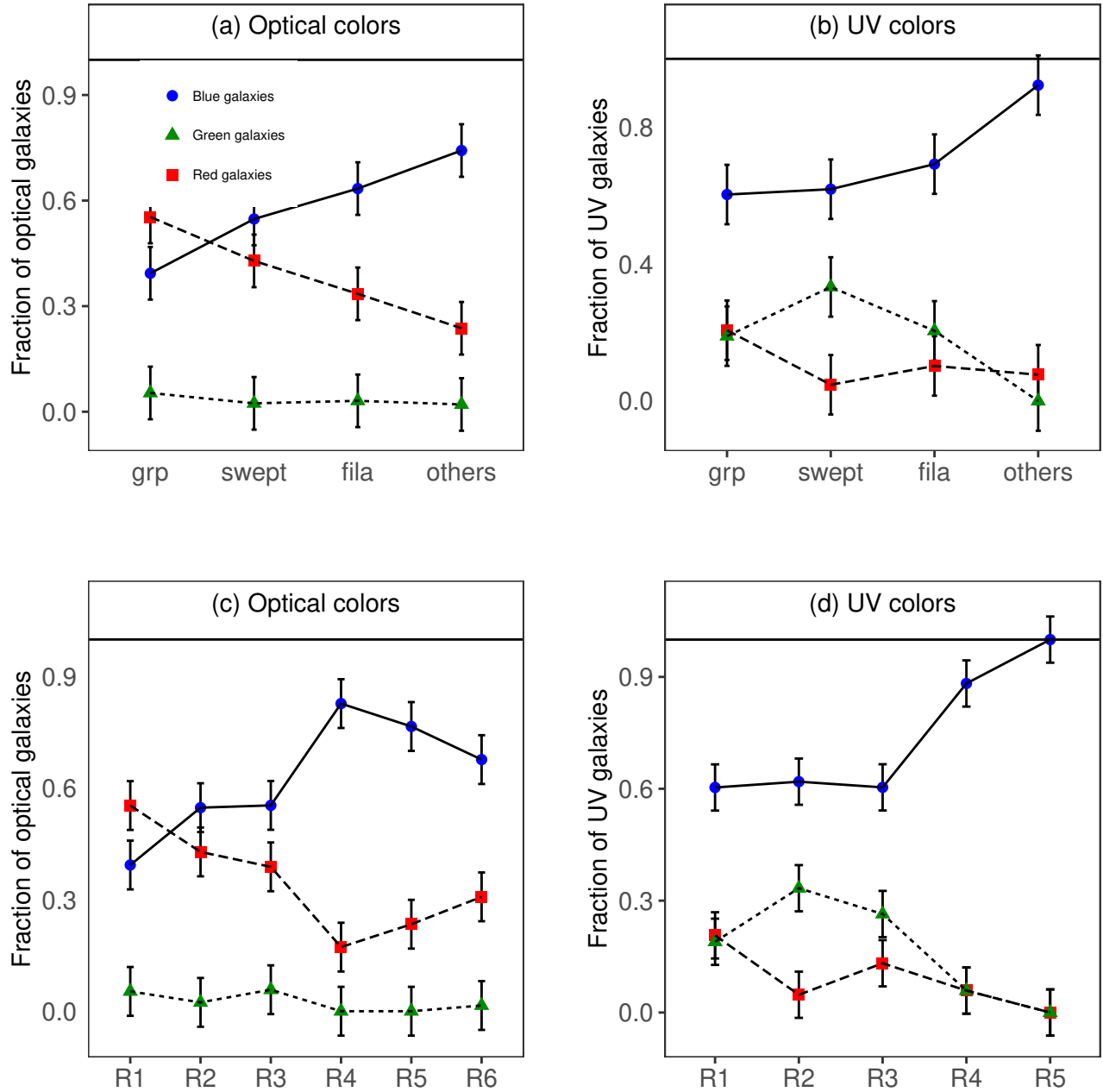} 
\caption{Fractional numbers of red, blue and green galaxies in optical and UV colors as a function of their underlying structure as described in \ref{subsec:ENV} and Table~\ref{tab:SUBSTRUCTURE} (top panels), and as a function of the regions as described in Table~\ref{tab:SUBSTRUCTURE2} in Abell 1882 (bottom panels). The fractional number of the optical and the NUV colors are shown in the left column and the right column, respectively.} 
\label{fig:frac_color_plots}
\end{figure*}

{\textit{Fig.~\ref{fig:frac_color_plots}a{\textit{--}d}}} show the optical and $NUV$ color evolution of galaxies as a function of the underlying large scale structures ($\textit{top panels}$), and the regions defined in Section \ref{subsec:ENV} (also see Table~\ref{tab:SUBSTRUCTURE}) and Table~\ref{tab:SUBSTRUCTURE2}($\textit{bottom panels}$). The fractional number of the optically red galaxies is greater than that of blue galaxies within the galaxy groups ($\sim$60$\%$ of the galaxies in the galaxy groups are optically red), and comparable within the `swept-up' region ({\textit{Fig.~\ref{fig:frac_color_plots}a}}).  The fractional number of optically blue galaxies is maximum within the galaxy population categorized as `others' ($\sim$75$\%$), i.e., the galaxies that do not lie within the groups, filaments or the `swept-up' region. 

Unlike its optical counterpart, the ratio of $NUV$ red galaxies to blue galaxies is not so clear, because the low mass, red galaxies are severely undercounted ({\textit{Fig.~\ref{fig:frac_color_plots}b}}). The $NUV$ red, massive galaxies do not show any appreciable trend. About $52\%$ of these $NUV$ massive, red galaxies detected by GALEX are within the galaxy groups (i.e., Region 1), and $\sim$30$\%$ in Region 3, and none detected in Region 5. However, the number of $NUV$ blue galaxies show a sharp decline from the galaxy population categorized as `others', to the galaxy population within the groups, `swept-up' region and the filaments. This strongly indicated that the underlying large scale structures have a role to play in the quenching of SF in the galaxies.


{\textit{Fig.~\ref{fig:frac_color_plots}c\text{--}d}} trace the optical and UV colors as a function of their radial distances from the assumed center of Abell 1882. At a projected radial distance $M/M_{\odot}{R}$ $>$ $1.4 Mpc$ from the assumed center of Abell 1882 (i.e., outside the groups and the `swept-up' region), the population is dominated by the optically blue galaxies ({\textit{Fig.~\ref{fig:frac_color_plots}b}}). The fractional number of the optically red galaxies has the highest value at the group environment, indicating an overall decrease in the SF within the galaxy group environment in the past 2$\text{--}$3 Gyrs. However, instead of progressively increasing outwards from the group environment, the fractional number of blue galaxies peaks at Region 4 ($\sim$90$\%$ of the galaxies). Region 4 has a lower $\Sigma$ ($\sim$7 galaxies/$Mpc^{2}$) than both the Region 3 ($\sim$29 galaxies/$Mpc^{2}$) and Region 5 ($\sim$17 galaxies/$Mpc^{2}$).

The fractional numbers of blue galaxies in Region 1, Region 2 and Region 3 are comparable. However, similar to the optical counterpart, the fractional number of the $NUV$ blue galaxies drastically increases in Region 4 which has much lower $\Sigma$ than Region 1, Region 2 and Region 3. The $NUV$ red colors trace the massive galaxies only. The fractional number of the massive, red galaxies is less in Region 5 compared to Region 1 (i.e., the groups).

They also indicate an increased SF in the low-density Region 4, and that the color-density relation is already in place in the very early stages of cluster formation.

The above results indicate that the galaxy color ties very strongly with local galaxy density, even in a complex structure like Abell 1882. We identify the projected radial distance from the assumed center as a second order evolutionary driver.

\subsection{Spatial locations of the Green valley Galaxies in $NUV$}\label{subsec:nuv green}

\begin{figure}
\centering 
\includegraphics[scale=0.5, trim={0 0 0 0}, clip=true]{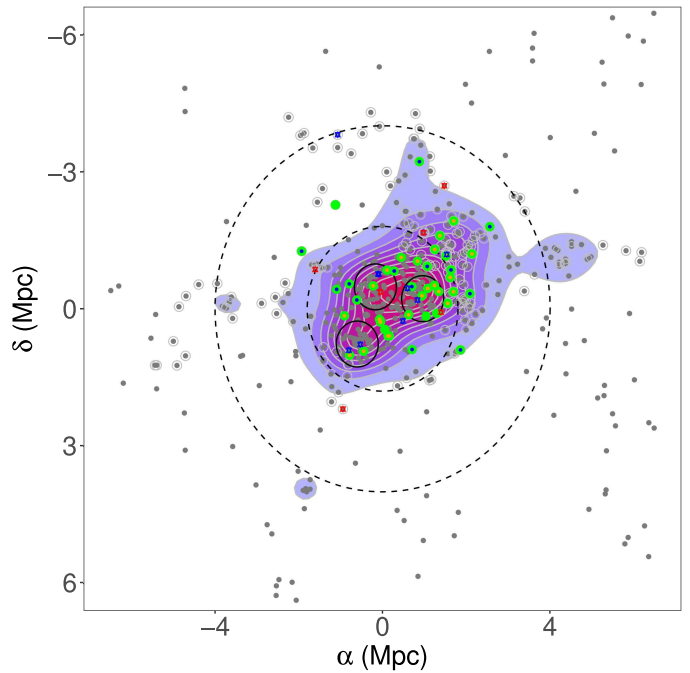} 
\caption{Local galaxy density map of Abell 1882 overlaid with Green Valley galaxies in $NUV$ (in green), red being the densest regions. The green galaxies with blue and orange dots indicate the galaxies that are optically blue and red, respectively. Plus signs in black indicate $H_{\alpha}$ emitting galaxies. Grey dots indicate all the galaxies in Abell 1882. Grey circles indicate the galaxies in the filament.} 
\label{fig:nuv_radec_bycolor_green2}
\end{figure}

The green dwarf galaxies may not be well represented due to the GALEX flux threshold. However, the fractional number of $NUV$ green galaxies increases inwards from Region 5 and peaks at the `swept-up' region (i.e., Region 2), indicating intermediate to high mass galaxies undergoing low level SF in the `swept-up' region, i.e., in the immediate infall region of the groups, in the past Gyr ({\textit{Fig.~\ref{fig:frac_color_plots}b, d}}). This population is conspicuous by its absence in the `others'.

{\textit{Fig.~\ref{fig:nuv_radec_bycolor_green2}}} shows the locations of the GV galaxies in the Supergroup (green filled circles), overlayed on a local galaxy density map. About $95\%$ of these GV galaxies lie within $\mathcal{R} < 2.6 Mpc$ from the assumed center of the SuperGroup, i.e., in Region 1, Region 2 and Region 3. About $27\%$ of the $NUV$ green galaxies lie within the galaxy groups (Region 1), and $\sim$68$\%$ of the $NUV$ galaxies lie in the immediate infall region of Abell 1882 ($\sim$34$\%$ each in Region 2 and Region 3). It is interesting to note that, although all the three outer regions, Region 3, Region 4 and Region 5, have a large fraction of galaxies in the filaments ($\sim$92$\%$, $\sim$81$\%$, and $\sim$53$\%$ galaxies in the filaments, respectively), the majority of the $NUV$ green galaxies that lie within the filaments ($\sim$80$\%$) are in Region 3 (radial distance 1.4 $<$ $\mathcal{R}$ $<$ 2.6 Mpc from the assumed center of Abell 1882). 
Hence, most of the low-level SF in the galaxies are close to the infall region, where the $\Sigma$ $\ge$ 29 $galaxies/Mpc^{2}$. Hence, eventhough there are GV galaxies within the galaxy group environment, the transitioning of the galaxies take place much before the galaxies are funneled into  the galaxy groups.

There are two distinct $NUV$ green galaxy populations shown in {\textit{Fig.~\ref{fig:nuv_radec_bycolor_green2}}}:

\noindent \textbf{{(i)}} \textit{The optically blue galaxies that are green in $NUV:$} The $NUV$ green galaxies (shown in green) with blue dots indicate the galaxies that are optically blue. These galaxies were actively star-forming in the past 2$\textit{--}$3 Gyrs. However, they have undergone only low-level SF in past 0.3$\textit{--}$1 Gyr, as indicated by their $NUV\textit{--}r$ color. About $30\%$ of the GV galaxies consists of this type of galaxies. Most of the galaxies in this population ($\sim$84$\%$) are in the immediate infall region of the groups (`swept-up' region, and galaxies inside the filaments close to the group environment; $\mathcal{R}$ $<$ 2.6 Mpc), indicating preprocessing in the high density infall region, which suppressed the high-level SF, and led to only low-level SF in the galaxies in the past 0.3$\textit{--}$1 Gyr. If we look at the mass distribution of the optically blue, $NUV$ green population, $\sim$46$\%$ are high mass galaxies, and $\sim$54$\%$ are intermediate mass galaxies.

\noindent \textbf{{(ii)}} \textit{The optically red galaxies that are green in $NUV:$} The $NUV$ green galaxies (shown in green) with orange dots indicate the galaxies that are optically red.  
About $60\%$ of the GV galaxies in $NUV$ are optically red. This galaxy population was passive between 2$\textit{--}$3 Gyrs. However, they have undergone low-level SF in past 0.3$\textit{--}$1 Gyr, as indicated by the green $NUV\textit{--}r$ color. This suggests recent SF or `rejuvenation' of the red galaxies (see \citealp{Yi2005, 2010MNRAS.404.1775T}) in at least the past Gyr. As discussed in Section \ref{subsec:CMD}, this rejuvenation can be caused due to minor, gas-rich mergers in passive galaxies (\citealp{2005MNRAS.360...60K}), or accretion of gas by the RS galaxies from the intergalactic medium (\citealp{2007ApJS..173..572T}).

All the galaxies in this population are in the immediate infall region of the groups (`swept-up' region, and galaxies inside the filaments close to the group environment; $\mathcal{R}$ $<$ 2.6 Mpc). All mass ranges show `rejuvenation' in the past 0.3$\textit{--}$1 Gyr. About $69\%$ of the intermediate mass galaxies, $33\%$ of the high mass galaxies, and $\sim$33$\%$ of the dwarf galaxies in the $NUV$ GV region are optically red. 

Hence, about $20\%$ of all high mass galaxies and $\sim$7$\%$ of all intermediate mass galaxies detected in $NUV$, that were actively star-forming in the past 2$\text{--}$3 Gyr, are currently undergoing only low-level SF. None of the dwarf galaxies in our sample are in this population, possibly because they are undercounted in $NUV$. On the other hand,  about $10\%$ of all the high mass galaxies and $\sim$20$\%$ of all the intermediate mass galaxies detected in $NUV$, that were passive in the past 2$\text{--}$3 Gyrs, are currently undergoing `rejuvenation'.

We also detect 5 optically blue high-mass galaxies that are red in $NUV$ (indicated by a red star), indicating quenching of these massive galaxies in the past 0.3$\textit{--}$1 Gyr. However, transformation of optically red galaxies to $NUV$ blue galaxies (indicated by blue star) are seen only in the lower mass galaxies ($log$ $M/M_{\odot}$ $<$ $10$). This means that the rejuvenation in the massive galaxies in the past 0.3$\textit{--}$1 Gyr are due to low-level SF, and not due to high-level continuous SF. These optically red, $NUV$ blue galaxies are mostly within the galaxy groups or very close to them.

\subsection{$D_{n}4000$ Index, $H_{\alpha}$, Optical Color and UV Color Histograms For Different Mass Bins} 
\label{subsec:D4000}

\begin{figure*}
\centering 
\includegraphics[scale= 1 , trim={0 9.3cm 0 0 }, clip=true]{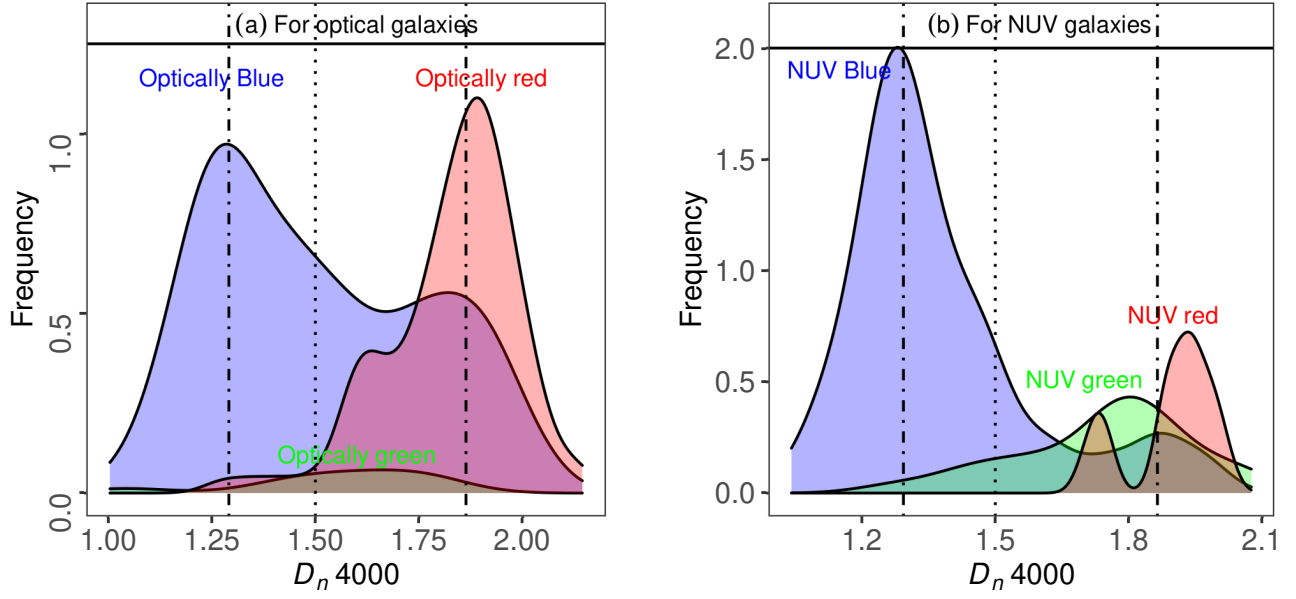} 
\caption{Smoothed histograms of $D_{n}4000$ index histograms for red (shown in red), blue (shown in blue) and green galaxies (shown in green) in (a) optical colors, (b) $NUV$ colors. The dashed and dotted lines represent the modes for the entire population in each category. The dotted line represents $D_{n}4000$ index $\sim$1.5 that corresponds to a stellar population of 1 Gyr.}
\label{fig:d4000_histograms}
\end{figure*}

 \begin{figure}
\centering 
\includegraphics[scale=0.5, trim={0 0 0 0 }, clip=true]{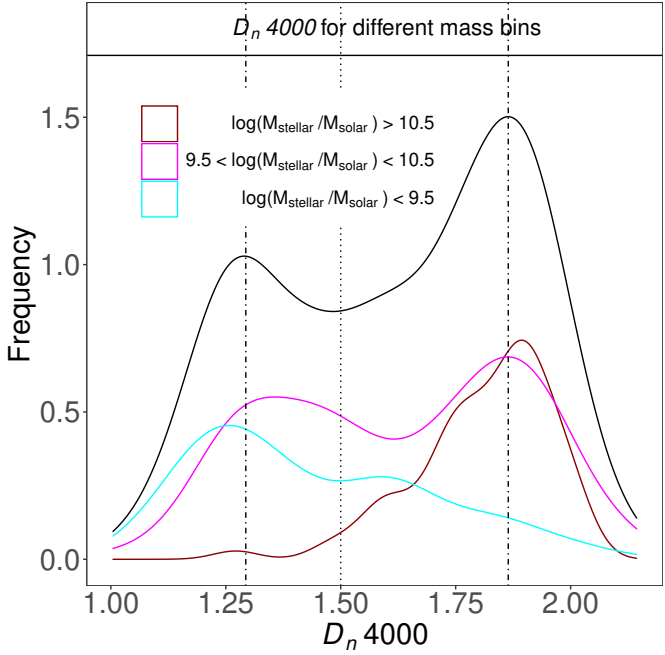} 
\caption{Smoothed histograms of $D_{n}4000$ index histograms for different mass bins. The black histogram represents all the galaxies. The dashed and dotted lines represent the modes for the entire population in each category. The dotted line represents $D_{n}4000$ index $\sim$1.5 that corresponds to a stellar population of 1 Gyr. Galaxy mass bins are defined in Section \ref{subsec:MASS}.} 

\label{fig:d4000_hist_mass}
\end{figure}

\begin{figure}
\centering 
\includegraphics[scale=0.5, trim={0 0 0 0}, clip=true]{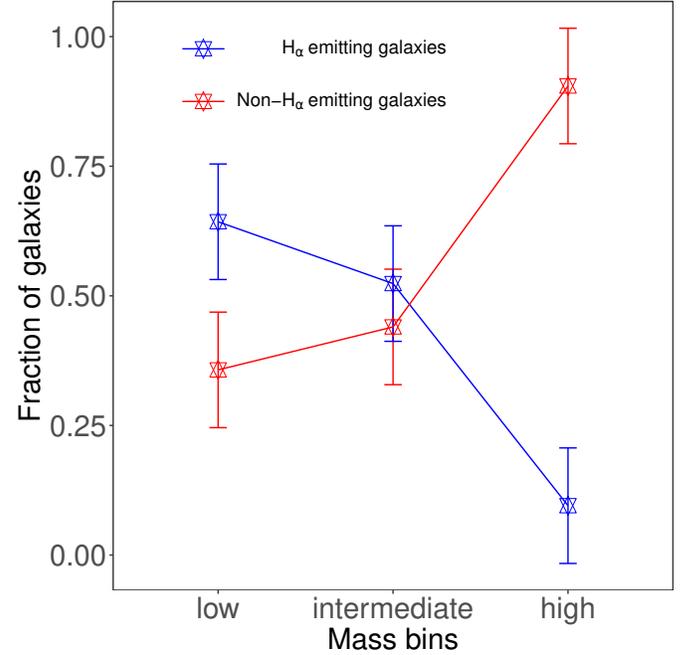} 
\caption{Fraction of $H_{\alpha}$ emitting (in blue) and non-$H_{\alpha}$ emitting galaxies (in blue) for different mass ranges. Galaxy mass bins are defined in Section \ref{subsec:MASS}.} 
\label{fig:frac_ewha_mass}
\end{figure}

The $D4000$ index or the 4000{\AA} break is the difference between the levels of continuum just blueward and redward of 4000{\AA}. In hot stars, multiply ionized elements reduce the opacity, and hence reduce the $D4000$ index. A strong $D4000$ index indicates a lack of hot, blue stars in the galaxy. Because of its smaller wavelength range, the $D4000$ index is less sensitive to dust attenuation of the stars, whose spectrum cover these wavelengths, compared to the broadband color parameters. (see \citealp{1983ApJ...273..105B, Hamilton, Hathi, Haines, 1999ApJ...527...54B, 2003MNRAS.341...54K}). 

The $D4000$ index is very sensitive to significant SF in the past 1 Gyr from the observed epoch, and also the metallicity. A $D4000\sim$1.5 indicates an average age of 1 Gyr for the stellar population, and this value can be used to separate star-forming and quiescent galaxies. (\citealp{2003MNRAS.341...54K, Hathi,  Haines}).

We have used a narrow version of the $D4000$ index, i.e., $D_{n}4000$ index as defined by \citet{1999ApJ...527...54B}. The $D_{n}4000$ index is much less sensitive to the reddening effect than the standard $D4000$. 

The stellar population is assumed to have $D_{n}4000\approx$1 at birth, and evolves redward \citep{1999ApJ...527...54B}. In addition, the $D_{n}4000$ index has a well defined upper limit of $\sim$2.3 for early type galaxies. Hence, we have retained  $D_{n}4000$ index values only within the range 1 $<$ $D_{n}4000$ $<$ 2.3 \citep{1983ApJ...273..105B, Hamilton, 2003MNRAS.341...54K,  kim}.

{\textit{Fig.~\ref{fig:d4000_histograms}a,b}} show the histograms of the $u\text{--}r$ and $NUV\text{--}r$ RS, GV, and BC galaxies in $D_{n}4000$ index bins. The modes for the overall population for all colors is shown in black dashed and dotted lines. $D_{n}4000\sim$1.5 is shown in black dotted line for reference. 

In {\textit{Fig.~\ref{fig:d4000_histograms}b}}, the modes of $D_{n}4000$ are well separated for red and blue galaxies in $NUV\text{--}r$ color as expected. We find that the modes of blue galaxies (shown with a blue histogram) and the red galaxies (shown with a red histogram) for both the $u\text{--}r$ and $NUV\text{--}r$ colors are at $D_{n}4000$ $\sim$1.29 and $D_{n}4000$ $\sim$1.86, respectively. This is very similar to the values obtained by \citet{2003MNRAS.341...54K}, who identified the peaks at $\sim$1.27 and $\sim$1.85, respectively. 
The $NUV$ blue galaxies represent a younger stellar population, whereas the red galaxies, which peak at $D_{n}4000$ $>$ 1.86, represent an older, metal-rich, mostly quiescent stellar population. A second, smaller peak in the $NUV\text{--}r$ red galaxies points at the possibility of two distinct quiescent stellar populations, where the smaller peak has a smaller mean stellar age. However, all the red galaxies have $D_{n}4000$ $>$ 1.5. This means that all the red $NUV$ galaxies have stopped forming stars at least 1 Gyr before the observed epoch. 

The $NUV\text{--}r$ GV galaxies peak at $D_{n}4000$ $>$ 1.5. It is important to note that the photometric colors are obtained from integrated light of the entire galaxy, whereas the spectral data are obtained only from the central region of the galaxies. Hence, the $NUV\text{--}r$ and $u\text{--}r$ colors may not represent the same environments as the regions within the galaxies represented by $D_{n}4000$. The $D_{n}4000$ obtained from SDSS gives us the stellar age over a 3$\arcsec$ diameter, whereas the $NUV$ data obtained from GALEX was selected within a  5$\arcsec$ radius of each optical galaxy. It is likely that the low-level SF detected in the $NUV$ GV galaxies are occurring in the outer regions of these galaxies, beyond the 3$\arcsec$ diameter. This result is similar to the evidence of low-level, outer disk star formation in the GALEX Nearby Galaxies Survey (NGS) (\citealp{2007ApJS..173..185G}. Also see \citealp{2007ApJS..173..572T}).

In {\textit{Fig.~\ref{fig:d4000_histograms}a}}, almost all the optically red galaxy population in the $u\text{--}r$ color have $D_{n}4000$ $>$ 1.5 as expected. The optically blue galaxy population in the $u\text{--}r$ color, has a significant number of galaxies with $D_{n}4000$ $>$ 1.5. This population of $u\text{--}r$, blue galaxies were star-forming in the past 2$\text{--}$3 Gyr from the observed epoch, but has been quiescent in the past 1 Gyr from the observed epoch, and contain mostly older, metal-rich galaxies.

{\textit{Fig.~\ref{fig:d4000_hist_mass}}} shows the histogram of $D_{n}4000$ index in mass bins. The peak values of $D_{n}4000$ are well separated for the high mass galaxies (shown in brown), and the dwarf galaxies (shown in cyan). This signifies that the dwarf galaxies mostly consist of younger stellar population, whereas the massive galaxies are older and metal-rich galaxies, and have stopped forming stars at least 1 Gyr before the observed epoch. The intermediate galaxies (shown in magenta) have comparable number of younger and older stellar populations. This is complemented by {\textit{Fig.~\ref{fig:frac_ewha_mass}}}, which shows that the dwarf galaxies have a higher fraction of $H_{\alpha}$ emitters ($65\%$), and the high mass galaxies are primarily non-$H_{\alpha}$ emitters ($90\%$). The intermediate mass galaxies have comparable numbers of $H_{\alpha}$ emitters and non-$H_{\alpha}$ emitters, where $H_{\alpha}$ emitters are currently forming stars. Hence, a separation of the galaxy population into mostly star-forming, dwarf galaxies (log $M_{\odot}$ $\leq$ 9.5), and quiescent, high-mass galaxies (log $M_{\odot}$ $\geq$ 10.5) have been in place in Abell 1882 for at least past 1 Gyr.

\subsection{Mass-Dependent Evolution Of $u\text{--}r$ And $NUV\text{--}r$ Colors In Different Environments Within Abell 1882} \label{subsec:MASS-EVOL}

\begin{figure*}[t]
\centering 
\includegraphics[scale=1, trim={0 0 0 0}, clip=true]{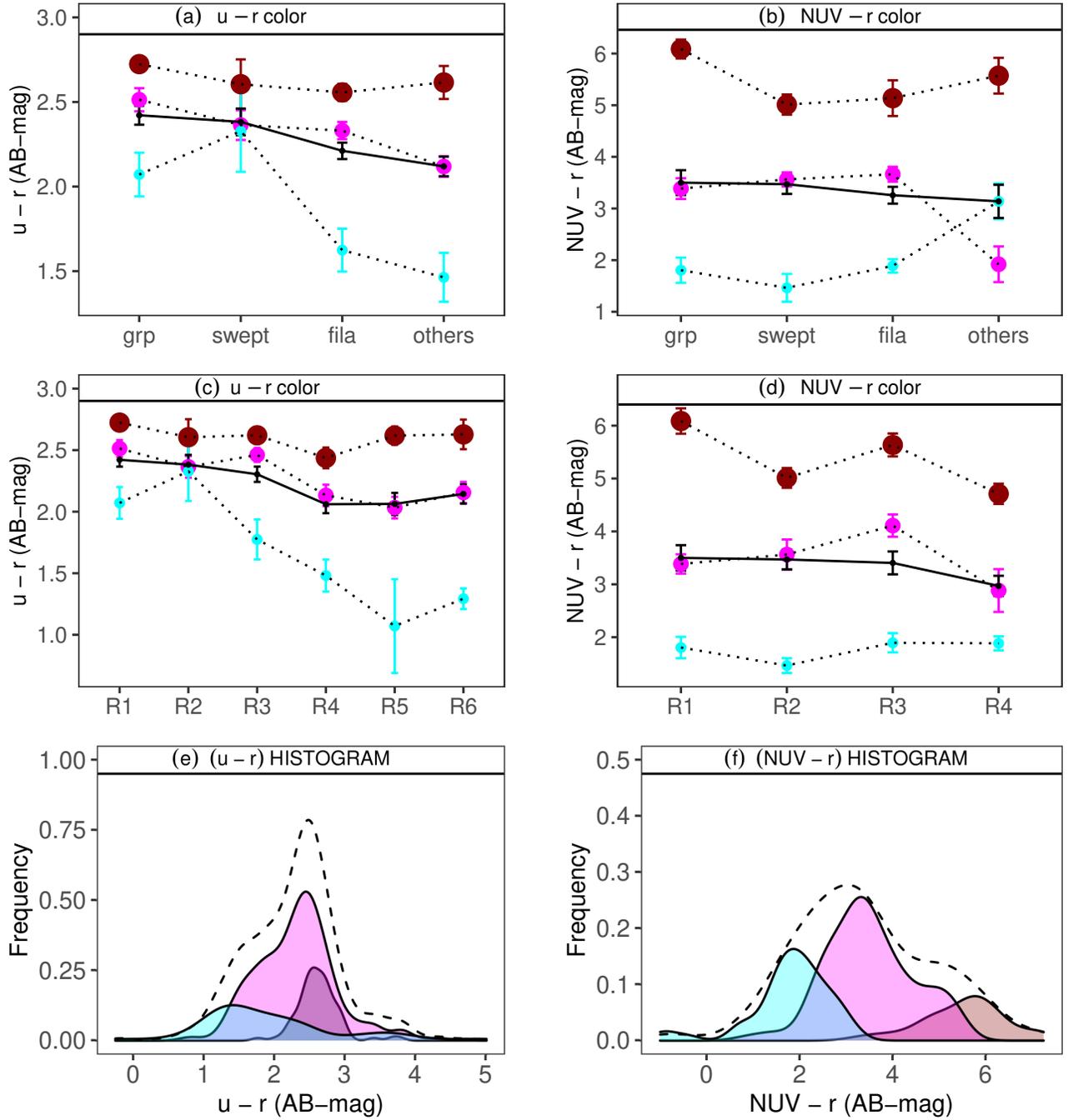}
\caption{\textbf{\textit{Top Panels:}} Mean $u\text{--}r$ color and $NUV\text{--}r$ color, respectively, with standard error bars, for galaxies of all masses (in black), high mass galaxies (in brown, large solid circles), intermediate mass galaxies (in magenta, smaller solid circles) and dwarf galaxies (in blue, smallest solid circles) ranges within the various underlying large scale structures of Abell 1882. The galaxy mass ranges are as defined in Section \ref{subsec:MASS}. \textbf{\textit{Middle Panels:}} Evolution of mean $u\text{--}r$ color and $NUV\text{--}r$ colors with radial distance form the assume core of Abell 1882. \textbf{\textit{Bottom Panels:}} Smoothed histograms of color distributions in dwarf galaxies (cyan), intermediate mass galaxies (magenta), and massive galaxies (brown). The dotted lines represent smoothed histogram of all galaxies for $u\text{--}r$ color and $NUV\text{--}r$ color, respectively.} 
\label{fig:mean_ur_nuvmass_plot}
\end{figure*}

In {\textit{Fig.~\ref{fig:mean_ur_nuvmass_plot}a$\text{--}$d}}, we trace the optical and UV color evolutions of the galaxies within the various underlying large scale structures of Abell 1882 $\textit{(top panels)}$, and within the different regions of Abell 1882 as defined in Section \ref{subsec:ENV} and Table~\ref{tab:SUBSTRUCTURE2} $\textit{(middle panels)}$, for different mass bins. The mass bins are defined in Section \ref{subsec:MASS}. The color evolutions are traced for all galaxies (in black solid line), as well as in three separate galaxy mass bins: high mass galaxies (in brown, large solid circles), intermediate mass galaxies (in magenta, smaller solid circles), and dwarf galaxies (in blue, smallest solid circles). {\textit{Fig.~\ref{fig:mean_ur_nuvmass_plot}a and c}}, show an overall reddening of the galaxies within the groups and the `swept-up' region in optical colors. 

This indicates an overall suppression of SF in the past $\sim$2$\text{--}$3 Gyrs. {\textit{Fig.~\ref{fig:mean_ur_nuvmass_plot}b}} shows no such trend in $NUV\text{--}r$ color within the different underlying large scale structures, although {\textit{Fig.~\ref{fig:mean_ur_nuvmass_plot}d}} shows a slight reddening of the $NUV\text{--}r$ color in Region 1, Region 2 and Region 3, compared to Region 4, highlighting the effect of radial distance and $\Sigma$ in quenching of SF in the galaxies. It is important to note that the majority of the galaxies in both the optical and UV colors are intermediate mass galaxies ($62\%$ of the galaxies in the optical colors, and $56\%$ of the galaxies in the NUV). Hence the mean color for all the galaxies in each bin is driven by the mean color of the intermediate galaxies in that bin.

The color evolutions of the galaxies in the optical and UV are more pronounced if we trace them in separate mass bins. The optical color samples the red, dwarf galaxies much more effectively than the $NUV$ color. Optically, the color evolution of the dwarf galaxies ($log$ $M/M_{\odot}$ $<$ $9.5$) is much more significant than the higher mass galaxies. The mean $u\text{--}r$ and $NUV\text{--}r$ colors for the high mass galaxies are the reddest, whereas that of the dwarf galaxies are the bluest, at all radial distances from the core, and for all underlying large scale structures. The colors of the galaxies in the individual mass bins are more separated in $NUV\text{--}r$ color, compared to the $u\text{--}r$ color.

{\textit{Fig.~\ref{fig:mean_ur_nuvmass_plot}c}} shows a drastic reddening of the dwarf galaxies in the optical color, as they approach the swept-up' region. This indicates quenching of dwarf galaxies in the `swept-up' region, i.e., in the immediate infall region of the group environment in the past 2$\text{--}$3 Gyrs. No such trend for the dwarf galaxies in observed in the $NUV\text{--}r$ color due to GALEX flux threshold for low luminosity galaxies that severely cuts the red and green dwarf galaxies in $NUV\text{--}r$ color. 

The intermediate mass galaxies within the groups, swept-up' region and the filaments are significantly redder in $NUV\text{--}r$ color, compared to the galaxies categorized as `others' ({\textit{Fig.~\ref{fig:mean_ur_nuvmass_plot}b}}). {\textit{Fig.~\ref{fig:mean_ur_nuvmass_plot}d}} also shows a reddening of the intermediate mass galaxies in Region 3 in $NUV\text{--}r$ color, indicating a decrease in SF in the intermediate mass galaxies at $\mathcal{R}$ $<$ 2.6 Mpc.

{\textit{Fig.~\ref{fig:mean_ur_nuvmass_plot}e$\text{--}$f}} show the smoothed histograms of $u\text{--}r$ color and $NUV\text{--}r$ color, respectively, shown in black dotted lines. The $NUV\text{--}r$ color histograms of high mass, intermediate mass and the dwarf galaxies are shown in dark grey, yellow and light grey colors, respectively. There is a clear bimodality in the optical and UV colors in the histograms for massive and dwarf galaxies. The $NUV\text{--}r$ color histogram for the intermediate mass galaxies peaks redward of the histogram of the dwarf galaxies, and blueward of the high mass galaxies. Compared to the UV color, the optical color-magnitude diagram in Abell 1882 does not show a well-separated RS and BC, especially at higher mass end.

\section{SUMMARY AND CONCLUSIONS} \label{sec:SUM}

We have amassed the largest optical catalog for Abell 1882 with 526 member galaxies using MMT Hectospec, the GAMA survey, the SDSS archive, NED and the Gemini Multi-Object Spectrographs. We have complemented the optical catalog with GALEX data, which has 191 $NUV$ detections. 
We have identified at least three primary feeding filaments in the cluster outskirts of Abell 1882, which accrete the galaxies asymmetrically mostly along the direction of the Right Ascension, through which most of the galaxies are eventually funneled into the groups. The local galaxy density $\Sigma$ increases sharply near the three major groups of galaxies. However, we observe distinct over-densities at various \rad in the SuperGroup outskirts, indicating that Abell 1882 is still in the process of growth and accretion. The infall pattern of the member galaxies in the SuperGroup has a characteristic trumpet shape that is usually observed in more massive and relaxed clusters like the Coma Cluster, even at this early stage of virialization. 

The average velocity dispersion in Abell 1882 is 620 km s$^{-1}$. This value is much lower than the velocity dispersion of massive and relaxed clusters like the Coma cluster, which has a velocity dispersion of $\sim$1000 km s$^{-1}$ (\citealp{1999ApJS..125...35S}). 
Filament 1 shows evidence of an undetected smaller galaxy group, or a galaxy group that is in the early stages of assembly. We have used optical and $NUV$ colors to determine the evolution of the galaxies as they are funneled into the galaxy groups at the center of Abell 1882. The over-abundance of Green Valley galaxies in $NUV\text{--}r$ color indicates that the UV colors are much better diagnostics of ongoing low-level SF compared to their optical counterpart. \\

Here are the questions we set out to address in this paper:\\

$\bullet$ At which point during the early evolutionary history of the formation of a cluster does one see significant galaxy transformations that lead to the over-abundance of optically red galaxies that are observed at the core of the present-day clusters? 
In other words, are the well established color-color and color-density relations seen in present-day clusters also seen in an unrelaxed cluster like Abell 1882? 

$\bullet$ Is there evidence for galaxy type transformation as a function of the local number density of the galaxies, and the spatial locations of the galaxies within the structure?\\

Here is a summary of the main results that address the above questions:\\

\noindent \textbf{\textit{$\bullet$(1)}} We observe that Abell 1882 has a well-formed Red Sequence, Green Valley, and Blue Cloud populations in their photometric colors, similar to more virialized clusters like Coma and Virgo. The higher mass galaxies in Abell 1882 peak at the redward side of the color scale, and they are more constrained in the $u\text{--}r$ color, compared to the dwarf galaxies. High mass, $NUV$ blue stars are conspicuous by their absence, indicating that most of the massive galaxies ($log$ $M/M_{\odot}$$>$ $10.5$) have already moved to the $NUV$ RS, and stopped forming stars in the past Gyr,

even before they enter the far outskirts of Abell 1882. Hence, evolutionary processes similar to more relaxed groups have already taken place in this unrelaxed cluster system. However, the RS in Abell 1882 is not as tightly constrained as more virialized clusters like Coma and Virgo, indicating a larger spread in the age of the Red Sequence galaxies in Abell 1882 compared to more well-formed clusters. The red galaxies in Abell 1882 are more likely formed due to stochastic episodes arising from various environmental mechanisms like mergers, ram-pressure stripping, harassments etc. spread over different epochs. The intrinsic scatter within the RS will likely reduce as the system evolves. 

\noindent \textbf{\textit{$\bullet$(2)}} We also find that, similar to the results found in SuperGroups 1120 -1202 \citep{2005ApJ...624L..73G,  2008ApJ...688L...5K}, there is already a higher fraction of red sequence galaxies ($60\%$ of optical RS, and $52\%$ of $NUV$ RS) in the individual groups in SuperGroup Abell 1882 compared to that in the far outskirts, indicating an overall decrease in the SF within the galaxy group environment at least in the past 2$\text{--}$3 Gyr.  The immediate infall region, i.e., the `swept-up' region also has a comparable fraction of red galaxies, indicating that the SF in the galaxies have been suppressed even before the galaxies enter the galaxy group environment. The overall decrease in the fraction of blue galaxies and increase in the fraction of red galaxies within the groups, filaments and the `swept-up' region indicate more preprocessing in the galaxies within the large scale structures (groups and filaments), and in the immediate infall region of the galaxies (`swept-up' region), compared to the galaxy population in `others', which contain a higher fraction of blue galaxies in both the optical and UV colors.

In the context of the projected radial distance from the assumed center of Abell 1882, we find that  at a projected radial distance $\mathcal{R}$ $>$ 1.4 Mpc from the assumed center of Abell 1882 (i.e., outside the groups and the `swept-up' region), the population is dominated by the optically blue galaxies.

However, Region 4, which is at a closer projected radial distance from the central group environment than Region 5, but has a much lower $\Sigma$ than Region 5, contains a higher fraction of optically blue galaxies and a lower fraction of optically red galaxies. This indicates an increased SF in the low-density Region 4 compared to Region 5, even though it is closer to the group environment. We identify $\Sigma$ as the primary, and the projected radial distance from the assumed center as a second order evolutionary driver.

Hence, similar to the findings of the past three decades of galaxy evolution in cluster environment, our results show that $\Sigma$, proximity to the central gravitational potential $\mathcal{R}$, as well as underlying LSS are important driving mechanisms for galaxy transformation, even within a complex, unvirialized cluster environment like Abell 1882. \\

We also constrained how the galaxy transformation is dependent on the mass of the galaxy.\\

\noindent \textbf{\textit{$\bullet$(3)}} We show that the dwarf galaxies (log $M_{\odot}$ $\leq$ 9.5) are  mostly star-forming, and that the high mass galaxies (log $M_{\odot}$ $\geq$ 10.5) are quiescent and metal-rich. The optical color evolution of the dwarf galaxies is much more pronounced than that of the higher mass galaxies. We see quenching of dwarf galaxies in the `swept-up' region ($\mathcal{R}$ $<$ 1.4 Mpc), i.e., in the immediate infall region of the group environment, in the past 2$\text{--}$3 Gyrs. The high mass galaxies, appear to have undergone most of the transformations, and are already on the Red Sequence  before they encounter the Abell 1882 environment. However, the massive galaxies are not all 'red and dead'. A significant number of these galaxies show transformation due to low-level SF (`rejuvenation'). Most of these transformations occur in the immediate infall region the group environment ($\mathcal{R} < 2.6 Mpc$). The intermediate mass galaxies also show an overall reddening of the galaxies at distances $\mathcal{R}$ $<$ 2.6 Mpc from the assumed center of Abell 1882. Further, it is likely that the low-level SF detected in the $NUV$ GV galaxies are primarily occurring in the outer regions of these galaxies, beyond the 3$\arcsec$ diameter, which is not sampled by the $D_{n}4000$.

To summarize, the optical color evolution (past 2$\text{--}$3 Gyrs) of the dwarf galaxies is most evident at roughly twice the virial radius of the groups, whereas the high and intermediate mass Green Valley galaxies indicate that the recent transformations in the past Gyr, within roughly four times the virial radius. Hence, filaments and the immediate infall region play a significant role in suppression of SF in the galaxies even before they enter the group environment.

We conclude that the physical mechanisms that suppress SF, primarily in the intermediate mass and dwarf galaxies, must be the ones that take effect in environments that have lower density than the galaxy groups, and hence occurs even before the galaxies fall into the high density, high velocity dispersion fields of the galaxy groups. This is similar to the findings of \citet{2008ApJ...685L.113S} and \citet{2009ApJ...705..809T}, who found evidence for quenching of SF in galaxies even before they enter the group environment. This strongly indicates low-density driven galaxy transformation mechanisms.

In hierarchical merger of galaxies, galaxy mergers can cause the combined dark matter halo to cross a critical halo mass ($>$$10^{12}$$M_{\odot}$), generating a shock in the accreting gas, as well as gas heating, and thus preventing it from cooling and forming stars, rendering them red and dead by redshift $z$ $\sim$ 0. If the galaxies reside in smaller halos, the gas may not be shock heated. This allows for the SF and subsequent regulation of the same by several feedback mechanisms. This model is favored by the hierarchical structure formation theory in the $\Lambda$CDM model. Combined with the fact that these massive galaxies form a tight, narrow red sequence, and are possibly primarily massive ellipticals, it strongly implies dry mergers as the driving force that causes the observed color of these galaxies. Dry mergers occur between already gas-poor, passive galaxies. Hence it doesn't lead to significant changes in the SF in these galaxies. This also explains why the color evolution of the massive galaxies is not as drastic as in the lower mass galaxies. However, at least $30\%$ of these massive galaxies show signs of rejuvenation close to the group environment. This indicates that at least some of these massive galaxies are undergoing low-level SF.

Dwarf galaxies and intermediate mass galaxies undergo significant transformation much before they enter the galaxy group environment. Filaments and the immediate infall region play a significant role in suppressing SF in these galaxies. We propose galaxy-galaxy wet mergers, which work in the low galaxy density and low velocity dispersion cases. We conclude that wet mergers may cause a more drastic change in the star-forming capabilities of the dwarf galaxies, which are primarily emission line galaxies, compared to the intermediate galaxies. This may lead to a more drastic color evolution and a lower $D_{n}4000$ index evolution in the dwarf galaxies compared to higher mass galaxies.

Since unrelaxed clusters at high redshifts are hard to detect, and are rare at lower redshifts, we have little understanding of the galaxy transformations occurring in these environments. However, the galaxy bimodality that we see in the present-day clusters trace back to the evolutionary mechanisms in the earlier phases of the cluster formation. This makes these unrelaxed cluster environments an inevitable part of the narrative in the study of galaxy evolution, in addition to telling us what happened in the past evolutionary history of a rich cluster like Coma. Our results for intermediate mass and dwarf galaxies are in agreement with that of SuperGroups Eridanus and SG1120-1202, which show an overabundance of early-type galaxies, indicating that morphological preprocessing similar to rich clusters has already taken place in these early systems, and possibly within the filaments (\citealp{2008ApJ...688L...5K, 2008galx.prop...18F, 2005ApJ...624L..73G}). We also find that the evolution of optical and $NUV$ colors is not only dependent on the $\Sigma$ and proximity to the central gravitational potential, but also on the galaxy mass \citep{2018MNRAS.474.4056M}.

\section {ACKNOWLEDGEMENTS}

Portions of this work were presented by Sengupta, A. to the University of Alabama in partial fulfillment of the requirement for the Ph.D. 

Observations reported here were obtained at the MMT Observatory, a joint
facility of the University of Arizona and the Smithsonian Institution.
We thank the MMT staff for their very skillful help in making the MMT
Hectospec observations.

Funding for the SDSS and SDSS-II has been provided by the Alfred P. Sloan Foundation, the Participating Institutions, the National Science Foundation, the U.S. Department of Energy, the National Aeronautics and Space Administration, the Japanese Monbukagakusho, the Max Planck Society, and the Higher Education Funding Council for England. The SDSS Web Site is http://www.sdss.org/.

The SDSS is managed by the Astrophysical Research Consortium for the Participating Institutions. The Participating Institutions are the American Museum of Natural History, Astrophysical Institute Potsdam, University of Basel, University of Cambridge, Case Western Reserve University, University of Chicago, Drexel University, Fermilab, the Institute for Advanced Study, the Japan Participation Group, Johns Hopkins University, the Joint Institute for Nuclear Astrophysics, the Kavli Institute for Particle Astrophysics and Cosmology, the Korean Scientist Group, the Chinese Academy of Sciences (LAMOST), Los Alamos National Laboratory, the Max-Planck-Institute for Astronomy (MPIA), the Max-Planck-Institute for Astrophysics (MPA), New Mexico State University, Ohio State University, University of Pittsburgh, University of Portsmouth, Princeton University, the United States Naval Observatory, and the University of Washington. 
GAMA is a joint European-Australasian project based around a spectroscopic campaign using the Anglo-Australian Telescope. The GAMA input catalogue is based on data taken from the Sloan Digital Sky Survey and the UKIRT Infrared Deep Sky Survey. Complementary imaging of the GAMA regions is being obtained by a number of independent survey programs including GALEX MIS, VST KiDS, VISTA VIKING, WISE, Herschel-ATLAS, GMRT and ASKAP providing UV to radio coverage. GAMA is funded by the STFC (UK), the ARC (Australia), the AAO, and the participating institutions. The GAMA website is http://www.gama-survey.org/.

We acknowledge the support from NASA JWST Interdisciplinary Scientist grants NAG5$-$12460, NNX14AN10G and 80NSSC18K0200 from GSFC.



\begin{thebibliography}{}



\bibitem[Baldry et al.(2004)]{2004ApJ...600..681B} Baldry, I.~K., Glazebrook, K., Brinkmann, J., et al.\ 2004, \apj, 600, 681 
\bibitem[Baldry et al.(2006)]{2006MNRAS.373..469B} Baldry, I.~K., Balogh, M.~L., Bower, R.~G., et al.\ 2006, \mnras, 373, 469 
\bibitem[Balogh et al.(2004)]{2004ApJ...615L.101B} Balogh, M.~L., Baldry, I.~K., Nichol, R., et al.\ 2004, \apjl, 615, L101 
\bibitem[Balogh et al.(1999)]{1999ApJ...527...54B} Balogh, M.~L., Morris, S.~L., Yee, H.~K.~C., Carlberg, R.~G., \& Ellingson, E. \ 1999, \apj, 527, 54 

\bibitem[Bell \&  de~Jong(2001)]{2001ApJ...550..212B} Bell, E.~F., de~Jong, R.~S. \ 2001, \apj, 550, 212-229



\bibitem[Berrier et al.(2009)]{2009ApJ...690.1292B} Berrier, J.~C., Stewart, K.~R., Bullock, J.~S., et al.\ 2009, \apj, 690, 1292 
\bibitem[Blanton et al.(2001)]{2001AJ....121.2358B} Blanton, M.~R., Dalcanton, J., Eisenstein, D., et al.\ 2001, \aj, 121, 2358 
\bibitem[Blanton et al.(2003)]{2003ApJ...594..186B} Blanton, M.~R., Hogg, D.~W., Bahcall, N.~A., et al.\ 2003, \apj, 594, 186

\bibitem[Brough et al.(2006)]{2006MNRAS.370.1223B} Brough, S., Forbes, D.~A., Kilborn, V.~A., \& Couch, W.\ 2006, \mnras, 370, 1223 

\bibitem[Bruzual(1983)]{1983ApJ...273..105B} Bruzual A.,~G.\ 1983, \apj, 273, 105


\bibitem[Cardelli et al.(1989)]{1989ApJ...345..245C} Cardelli, J.~A., Clayton, G.~C., \& Mathis, J.~S.\ 1989, \apj, 345, 245 
\bibitem[Colless et al.(2001)]{2001MNRAS.328.1039C} Colless, M., Dalton, G., Maddox, S., et al.\ 2001, \mnras, 328, 1039 
\bibitem[Cooper et al.(2007)]{2007MNRAS.376.1445C} Cooper, M.~C., Newman, J.~A., Coil, A.~L., et al.\ 2007, \mnras, 376, 1445
\bibitem[Cucciati et al.(2006)]{2006astro.ph.12120C} Cucciati, O., Iovino, A., Marinoni, C., et al.\ 2006, arXiv:astro-ph/0612120 


\bibitem[Dressler(1980)]{1980ApJ...236..351D} Dressler, A.\ 1980, \apj, 236, 351 
\bibitem[Dressler et al.(1997)]{1997ApJ...490..577D} Dressler, A., Oemler, A., Jr., Couch, W.~J., et al.\ 1997, \apj, 490, 577 
\bibitem[Dressler et al.(1999)]{1999ApJS..122...51D} Dressler, A., Smail, I., Poggianti, B. M., Butcher, H., Couch, W. J., Ellis, R. S., Oemler, A., Jr.\ 1999, \apj, 122, 51


\bibitem[Erkurt et al.(2009)]{2009POBeo..86..305E} Erkurt, A., Tektunali, H.~G., Hudaverdi, M., \& Ercan, E.~N.\ 2009, Publications de l'Observatoire Astronomique de Beograd, 86, 305
\bibitem[Evans (1996)] {Evans} Evans, J.D., Straightforward Statistics for the Behavioral Sciences. Brooks/Cole Publishing; Pacific Grove, Calif.: 1996. 

\bibitem[Fabricant et al.(2005)]{2005PASP..117.1411F} Fabricant, D., Fata, R., Roll, J., et al.\ 2005, \pasp, 117, 1411 
\bibitem[Fabricant et al.(2008)]{2008PASP..120.1222F} Fabricant, D.~G., Kurtz, M.~J., Geller, M.~J., et al.\ 2008, \pasp, 120, 1222
\bibitem[Fadda(2008)]{2008galx.prop...18F} Fadda, D.\ 2008, GALEX Proposal, 18 
\bibitem[Fassbender et al.(2011)]{2011A&A...527L..10F} Fassbender, R., Nastasi, A., B{\"o}hringer, H., et al.\ 2011, \aap, 527, L10 

\bibitem[Martin et al.(2005)]{2005ApJ...619L...1M} Martin, D.~Christopher, Fanson, James, et al.\ 2005, \apjl, 619L, 1M

\bibitem[Morrissey et al.(2007)]{2007ApJS..173..682M} Morrissey, Patrick, Conrow, Tim, et al.\ 2007,  \apjs, 173, 682M
 
 

\bibitem[Liske et al.(2015)]{2015MNRAS.452.2087L} Liske, J., Baldry, I.~K., \ 2015, \mnras, 452, 2087L
 
\bibitem[Geller et al.(1999)]{1999ApJ...517L..23G} Geller, M.~J., Diaferio, A., \& Kurtz, M.~J.\ 1999, \apjl, 517, L23 
\bibitem[G{\'o}mez et al.(2003)]{2003ApJ...584..210G} G{\'o}mez, P.~L., Nichol, R.~C., Miller, C.~J., et al.\ 2003, \apj, 584, 210
\bibitem[Gomez et al.(2010)]{2010AAS...21535103G} Gomez, P.~L., Miller, C.~J., Ingraham, P.~J., \& Sifon, C.\ 2010, BAAS, 42, 351.03 
\bibitem[Gonzalez et al.(2005)]{2005ApJ...624L..73G} Gonzalez, A.~H., Tran, K.-V.~H., Conbere, M.~N., \& Zaritsky, D.\ 2005, \apjl, 624, L73 


\bibitem[Haines(2017)]{Haines} Haines, C.~P., Iovino, A., Krywult, J., et al. \ 2017, \aap, 605,  A4
\bibitem[Hamilton(1985)]{Hamilton} Hamilton, D. \ 1985, \apj, 297, 371


\bibitem[Hashimoto et al.(1998)]{1998ApJ...499..589H} Hashimoto, Y., Oemler, A., Jr., Lin, H., \& Tucker, D.~L.\ 1998, \apj, 499, 589 


\bibitem[Hathi(2009)]{Hathi} Hathi, N.~P., Ferreras, I., Pasquali, A., et al. \ 2009
\apj, 690, 1866


\bibitem[Huchra \& Geller(1982)]{1982ApJ...257..423H} Huchra, J.~P., \& Geller, M.~J.\ 1982, \apj, 257, 423

\bibitem[Gil de Paz et al.(2007)]{2007ApJS..173..185G} Gil~de~Paz, A., Boissier, S., Madore, B.~F. et al \apjs, 173, 185


\bibitem[Kauffmann et al.(2003)]{2003MNRAS.341...54K} Kauffmann, G., Heckman, T.~M., White, S.~D.~M., et al.\ 2003, \mnras, 341, 54
\bibitem[Kauffmann et al.(2004)]{2004MNRAS.353..713K} Kauffmann, G., White, S.~D.~M., Heckman, T.~M., et al.\ 2004, \mnras, 353, 713
\bibitem[Kautsch et al.(2008)]{2008ApJ...688L...5K} Kautsch, S.~J., Gonzalez, A.~H., Soto, C.~A., et al.\ 2008, \apjl, 688, L5


 
 
\bibitem[Kaviraj, et al.(2005)]{2005MNRAS.360...60K} Kaviraj, S., Devriendt, J.~E.~G., Ferreras, I., Yi, S.~K.\ 2005, \mnras, 360, 60

\bibitem[Kaviraj, et al.(2007)]{2007ApJS..173..619K} Kaviraj, S., Schawinski, K., et al. \ 2007,  \apjs, 173, 619K


\bibitem[Kennicutt(2004)]{2004yCat.7141....0K} Kennicutt, R.~C., Jr.\ 2004, VizieR Online Data Catalog, 7141

\bibitem[Kim et al.(2018)]{kim} Kim, K., Malhotra, S., Rhoads, J. E. et al. \ 2018 , \apj, 867, 118

\bibitem[Lewis et al.(2002)]{2002MNRAS.334..673L} Lewis, I., Balogh, M., De Propris, R., et al.\ 2002, \mnras, 334, 673 


\bibitem[Mahajan et al.(2010)]{2010MNRAS.404.1745M} Mahajan, S., Haines, C.~P., \& Raychaudhury, S.\ 2010, \mnras, 404, 1745 
\bibitem[Mahajan et al.(2011)]{2011MNRAS.412.1098M} Mahajan, S., Haines, C.~P., \& Raychaudhury, S.\ 2011, \mnras, 412, 1098

\bibitem[Martin et al.(2007)]{2007ApJS..173..342M} Martin, D.~C.; Wyder, T.~K.; Schiminovich, D.\ 2007, \apjs, 173, 342


\bibitem[McGee et al.(2009)]{2009MNRAS.400..937M} McGee, S.~L., Balogh, M.~L., Bower, R.~G., Font, A.~S., \& McCarthy, I.~G.\ 2009, \mnras, 400, 937 
\bibitem[Miller et al.(2010)]{2010AAS...21641801M} Miller, C.~J., Gomez, P.~L., Sifon, C.~A., et al.\ 2010, BAAS, 41, 418.01
\bibitem[McKinley et al.(2018)]{2018MNRAS.474.4056M} McKinley, B., Tingay, S.~J., Carretti, E., et al.\ 2018, \mnras, 474, 4056 
\bibitem[Morrison et al.(2003)]{2003ApJS..146..267M} Morrison, G.~E., Owen, F.~N., et al.\ 2003, \apjs, 146, 267


\bibitem[Owers et al.(2013)]{2013ApJ...772..104O} Owers, M.~S., Baldry, I.~K., Bauer, A.~E., et al.\ 2013, \apj, 772, 104


\bibitem[Pimbblet et al.(2002)]{2002MNRAS.331..333P} Pimbblet, K.~A., Smail, I., Kodama, T., et al.\ 2002, \mnras, 331, 333 

\bibitem[Planck Collaboration(2018)] {2018arXiv180706209P}  Planck Collaboration, Aghanim, N., et. al. \ 2018,  arXiv:1807.06209



\bibitem[Poggianti et al.(2006)]{2006ApJ...642..188P} Poggianti, B.~M., von der Linden, A., De Lucia, G., et al.\ 2006, \apj, 642, 188
\bibitem[Postman \& Geller(1984)]{1984ApJ...281...95P} Postman, M., \& Geller, M.~J.\ 1984, \apj, 281, 95 
\bibitem[Postman et al.(2005)]{2005ApJ...623..721P} Postman, M., Franx, M., Cross, N.~J.~G., et al.\ 2005, \apj, 623, 721


\bibitem[Rood(1970)]{1970ApJ...162..333R} Rood, H.~J.\ 1970, \apj, 162, 333

\bibitem[Alam et al.(2015)]{2015ApJS..219...12A} Alam, Shadab, Albareti, Franco D., \ 2015, \apjs, 219, 12A

\bibitem[Saintonge et al.(2008)]{2008ApJ...685L.113S} Saintonge, A., Tran, K.-V.~H., \& Holden, B.~P.\ 2008, \apjl, 685, L113 

\bibitem[Salim et al.(2007)]{Salim07} Salim, S., Rich, R.~M., Charlot, S., et al. \ 2007, \apjs, 173,267

\bibitem[Salim(2014)]{2014SerAJ.189....1S} Salim, S.\ 2014, Serbian Astronomical Journal, 189, 1
\bibitem[Smail et al.(1997)]{1997ApJS..110..213S} Smail, I., Dressler, A., Couch, W.~J., et al.\ 1997, \apjs, 110, 213


\bibitem[Spergel et al.(2003)]{2003ApJS..148..175S} Spergel, D.~N., Verde, L., Peiris, H.~V., et al.\ 2003, \apjs, 148, 175 
\bibitem[Springel et al.(2006)]{2006Natur.440.1137S} Springel, V., Frenk, C.~S., \& White, S.~D.~M.\ 2006, \nat, 440, 1137
\bibitem[Smith et al.(2005)]{2005ApJ...620...78S} Smith, G.~P., Treu, T., Ellis, R.~S., Moran, S.~M., \& Dressler, A.\ 2005, \apj, 620, 78
\bibitem[Spitzer \& Baade(1951)]{1951ApJ...113..413S} Spitzer, L.~Jr., \& Baade, W.\ 1951, \apj, 113, 413
\bibitem[Schawinski et al.(2014)]{2014MNRAS.440..889S} Schawinski, Kevin, Urry, C.~Megan et al., \ 2014,  \mnras, 591, 53T

\bibitem[Struble \& Rood(1999)]{1999ApJS..125...35S} Struble, M.~F., Rood, H.~J. \ 1999, \apjs, 125, 35


\bibitem[Thilker et al.(2007)]{2007ApJS..173..572T} Thilker, D.~A., Boissier, S., Bianchi, L. et al. \ 2007, \apjs, 173, 572

\bibitem[Thomas et al.(2010)]{2010MNRAS.404.1775T} Thomas, D., et al.\ 2010, \mnras, 404, 1775


\bibitem[Tran et al.(2009)]{2009ApJ...705..809T} Tran, K.-V.~H., Saintonge, A., Moustakas, J., et al.\ 2009, \apj, 705, 809

\bibitem[Treu et al.(2003)]{2003ApJ...591...53T} Treu, T., Ellis, R.~S., Kneib, J.~P., et al.\ 2003, \apj, 591, 53

\bibitem[Vijayaraghavan \& Ricker(2013)]{Vijay} Vijayaraghavan, R., Ricker, P.~M.\ 2013, \mnras, 435, 2713

\bibitem[Wyder et al.(2007)]{2007ApJS..173..293W} Wyder, T.~K., Martin, D.~C., Schiminovich, D. et al.\ 2007 \apjs, 173, 293


\bibitem[Yi et al.(2005)]{Yi2005} Yi S.~K. et al.\ 2005, \apjl, 619, L111

\bibitem[York et al.(2000)]{2000AJ....120.1579Y} York, D.~G., Adelman, J., Anderson, J.~E., Jr., et al.\ 2000, \aj, 120, 1579

\end{thebibliography}
\end{document}